\documentclass[a4paper,aps,prb,citeautoscript,showpacs,floatfix,nofootinbib,preprint]{revtex4}
\usepackage{mathptmx}
\usepackage[T1]{fontenc}
\usepackage[latin9]{inputenc}
\usepackage{xcolor}
\usepackage{pdfcolmk}
\usepackage{amsmath}
\usepackage{amssymb}
\usepackage{mathdots}
\usepackage{esint}
\usepackage[unicode=true, bookmarks=true, bookmarksnumbered=false, bookmarksopen=false, breaklinks=true, pdfborder={0 0 0}, backref=false, colorlinks=true]{hyperref}
\hypersetup{pdfauthor={Cristian A. Marocico}}
\hypersetup{pdftitle={A Theoretical Investigation of Decay and Energy Transfer Rates and Efficiencies Near Gold Nanospheres}}

\usepackage[caption=false]{subfig}
\usepackage{siunitx}
\usepackage[english]{babel}
\usepackage[]{graphicx}

\begin{document}

\preprint{Version 1.0 \today}
\title{A Theoretical Investigation of Decay and Energy Transfer Rates and Efficiencies Near Gold Nanospheres}

\author{Cristian A. Marocico}
\affiliation{Semiconductor Photonics Group, School of Physics and CRANN,\\
Trinity College Dublin, College Green 2, Dublin, Ireland}

\author{Xia Zhang}
\affiliation{Semiconductor Photonics Group, School of Physics and CRANN,\\
Trinity College Dublin, College Green 2, Dublin, Ireland}

\author{A. Louise Bradley}
\email{bradlel@tcd.ie}
\affiliation{Semiconductor Photonics Group, School of Physics and CRANN,\\
Trinity College Dublin, College Green 2, Dublin, Ireland}

\date{\today}
\begin{abstract}
We consider the effect of gold nanospheres of subwavelength sizes on the decay and energy transfer rates of quantum systems placed in the proximity of these nanospheres. We find that, for the sphere sizes considered in this contribution, the radiative decay rate is barely affected by the presence of the nanosphere, whereas the non-radiative decay rate is greatly enhanced due to energy transfer from the quantum system to the nanosphere, leading to a strong quenching of the emission of the quantum system. The emission wavelength of the quantum emitter and its intrinsic quantum yield play an important role and the impact of both has to be considered together when investigating their effect on the non-radiative decay rate. The energy transfer process from the emitter to the nanosphere presents a complicated distance dependence, with a $r^{-6}$ regime, characteristic of the F{\"o}rster energy transfer mechanism, but also exhibiting other distance dependence regimes. In the case of a donor-acceptor pair of quantum systems in the presence of a gold nanosphere, the donor couples strongly to the nanosphere, acting as an enhanced dipole; the donor-acceptor energy transfer rate then follows a F{\"o}rster trend, with an increased F{\"o}rster radius. The coupling of the acceptor to the nanosphere has a different distance dependence, and it does not follow a F{\"o}rster-type trend. The angular dependence of the energy transfer efficiency between donor and acceptor has a strong dipole-dipole trend for small spheres and deviating from it for larger spheres, especially when the donor and acceptor are on opposite sides of the sphere. The spectral overlap of the donor emission, acceptor absorption and gold nanosphere extinction/scattering shows an interesting trend in that the largest F{\"o}rster radius is obtained when the donor emission and acceptor absorption maxima are somewhat red-shifted from the localized surface plasmon peak in the extinction spectrum of the gold nanosphere, being located between it and the near-field scattering maximum.

\end{abstract}
\maketitle

\section{Introduction}
\label{sec:I}

The optical properties of quantum emitters (QEs), such as quantum dots and fluorescent dyes, near metal nanoparticles (NPs) are a topic of great interest both from the point of view of a fundamental understanding of these properties,\cite{Yun2005, Jennings2006, Singh2010} as well as the technological applications that they can lead to, e.g.~solar cells,\cite{Markvart2000, Catchpole2008, Beck2009, Mokkapati2009, Atwater2010, Jung2011, Mavrokefalos2012} light-emitting diodes,\cite{Achermann2006, Demir2008, Nizamoglu2008a, Demir2011, Demir2011a, Diedenhofen2011a, Seitz2012, Erdem2012} subwavelength imaging,\cite{Liao2012, Sapienza2012} micro-lasers,\cite{John1999, Woldeyohannes1999} single photon sources,\cite{Angelakis2004, Grzela2012, Jacob2012, Diedenhofen2011a} etc. 

Of particular importance is the influence that the \emph{localized surface plasmon} (LSP) excited at the surface of the metal nanoparticles\cite{Averitt1997, Berciaud2005, Ancey2007, Ancey2009, Davis2010, Anghinolfi2012} has on the optical properties of quantum emitters. The influence of the LSP on the non-radiative decay rates of quantum emitters placed close to a single metal sphere has been extensively studied, both experimentally and theoretically.\cite{Yun2005, Jennings2006, Singh2010, Ruppin1982, Dung2000, Dung2001, Carminati2006, Moroz2010, Zhang2012, Sukharev2014} Different models have been proposed for the mechanism of the energy transfer process between the quantum emitter and the metal NP, such as the \textit{Nanometal Surface Energy Transfer} (NSET) model,\cite{Yun2005, Jennings2006, Singh2010} which considers the sphere as an infinite surface and has a distance dependence of the form $d^{-4}$, and the \textit{F{\"o}rster Resonance Energy Transfer} (FRET) model, which considers the sphere a polarizable dipole, and has a distance dependence of the form $r^{-6}$.\cite{Carminati2006} Recent theoretical work has shown that the distance dependence is more complex, and the NSET and FRET models are only valid in certain distance regimes.\cite{Sukharev2014}

The metal NP is an extended object and therefore has to be modeled in a full quantum electrodynamic formalism. When such a formalism is employed, the distance dependence of the non-radiative decay rate of the QE becomes more complex, including $r^{-6}$, $r^{-4}$, as well as other contributions, in agreement with previous theoretical work.\cite{Ruppin1982, Moroz2005, Mertens2007a, Moroz2010} In addition to the distance dependence, we investigate the effect of the emission wavelength and quantum yield, as well as the size of the metal NP on the energy transfer rate between the QE and the metal NP.

Of perhaps even more importance is the process of excitation energy transfer in a donor-acceptor pair, one of the main pathways by which energy transfer occurs at the nanoscale. It plays important roles in biology,\cite{Clegg1995,Andrews1999,Lai2013,Chen2014c,Holmstrom2014} nanophotonics (LEDs, nanolasers), microscopy,\cite{Zhang2008a, DeAngelis2013} sensing,\cite{Amjadi2014} and as optical rulers,\cite{Sonnichsen2005,Chatterjee2011,Samanta2014} etc. The process is determined by the dipole-dipole interaction between the donor and acceptor and, in the short distance limit, has a $r^{-6}$ dependence on the donor-acceptor separation $r$, the above-mentioned FRET regime.\cite{Forster1948}

The FRET regime is known to hold for donor-acceptor separations smaller than approximately $10\;\text{nm}$; for separations larger than this, the energy transfer process is overwhelmed by other de-excitation mechanisms of the donor, such as phonon relaxation or emission into the far-field. In a variety of applications, it is desirable to extend the FRET regime to larger donor-acceptor separations. This would enhance the efficiency of light-harvesting\cite{Law2005, Muskens2008} and emitting systems,\cite{Chanyawadee2010,Yoo2013} as well as increase the range and accuracy of sensing devices \cite{Amjadi2014} and optical rulers.\cite{Sonnichsen2005,Chatterjee2011,Samanta2014}

One possible way of enhancing the energy transfer rate in a donor-acceptor pair is by introducing a ``mediator''. The role of the mediator is to increase the probability that energy transfer between donor and acceptor will occur. One such mediator is a metal nanoparticle which can support LSPs, whose main characteristics are both a large confinement and a large enhancement of the electromagnetic field. The field of a LSP can thus provide a good mediator for the energy transfer process between donor and acceptor, when these are suitably positioned around the nanoparticle.

The field of the LSP on the metal nanoparticle influences not only the energy transfer rate between donor and acceptor, but also the other channels which contribute to the total decay rate of the donor and introduce new ones, such as absorption of energy by the metal. It is therefore important to enhance not only the energy transfer rate, but in particular the energy transfer efficiency, which is a measure of the competition between the donor-acceptor energy transfer rate and all other decay channels of the donor.

In this contribution we undertake an in-depth investigation of the role of the LSP on small spherical Au nanoparticles (Au NPs) in modifying the decay and energy transfer rates and efficiencies of quantum systems placed in the proximity of the Au NP. We begin in Sec.~\ref{sec:II} by laying out briefly the theoretical framework used in our investigations. Section \ref{sub:III.A} deals with the decay rate of a quantum emitter. In Sec.~\ref{subsub:III.A.1} we validate the model used by simulating experimental results obtained in our lab.\cite{Zhang2012} We then continue with an investigation of the decay rates of a single quantum emitter placed near a Au NP and the dependence of these rates on the emission wavelength and intrinsic quantum yield of the emitter (Sec.~\ref{subsub:III.A.2}), on the distance from the emitter to the Au NP (Sec.~\ref{subsub:III.A.3}) and on the size of the Au NP (Sec.~\ref{subsub:III.A.4}). In Sec.~\ref{sub:II.B} we consider the energy transfer rate between a donor-acceptor pair placed near the Au NP, beginning in Sec.~\ref{subsub:III.B.1} with a validation of the model through simulating experimental results obtained in our lab,\cite{Zhang2014} and continuing with an investigation of the dependence of the energy transfer efficiency in the case of a single donor-Au NP-acceptor triad on the donor and acceptor distance to the Au NP (Sec.~\ref{subsub:III.B.2}), on the relative angular position of the donor and acceptor around the Au NP (Sec.~\ref{subsub:III.B.3}), on the spectral overlap between donor emission, acceptor absorption and Au NP extinction spectra (Sec.~\ref{subsub:III.B.4}) and on the size of the Au NP (Sec.~\ref{subsub:III.B.5}). Finally, the conclusions are discussed in Sec.~\ref{sec:IV}.

\section{Theoretical Framework}
\label{sec:II}

Using a Green's tensor formalism, which we will briefly sketch in what follows, we can calculate the energy transfer rate and efficiency between two point dipoles placed near a metal nanoparticle.

\subsection{Decay Rates}
\label{sub:II.A}

Considering a quantum system modeled as a electric point dipole, its decay rate in the presence of the Au NP can be related to the power emitted as
\begin{equation}
	\label{eq:01}
	\frac{\gamma(\mathbf{r}, \omega)}{\gamma_0(\omega)} = \frac{P(\mathbf{r}, \omega)}{P_0(\omega)},
\end{equation}
where $\gamma(\mathbf{r}, \omega)$ and $\gamma_0(\omega)$ are the decay rates of the quantum system in the presence and absence of the Au NP, respectively, and $P(\mathbf{r}, \omega)$ and $P_0(\omega)$ are the power emitted by the dipole in the presence and absence of the Au NP, respectively.

The emitted power in the absence of the Au NP, $P_0(\omega)$ has the classical expression:\cite{Novotny2012}
\begin{equation}
	\label{eq:02}
	P_0(\omega) = \frac{n \omega^4 \mu_0^2}{12\pi\varepsilon_0 c^3}.
\end{equation}
The emitted power can be calculated from the integral of the normal component of the Poynting vector along a closed surface containing the quantum system:
\begin{equation}
	\label{eq:03}
	P(\mathbf{r}_0, \omega) = \frac{1}{2}\text{Re} \oint \textrm{d}\Omega\,\mathbf{\hat{n}} \cdot \mathbf{E}(\mathbf{r}, \omega) \times \mathbf{H}^*(\mathbf{r}, \omega).
\end{equation}
The total and radiative emitted power can be calculated from this expression by letting the imaginary surface shrink to a point or be infinite. The total power emitted in the presence of the sphere is
\begin{align}
	\label{eq:04}
	P(\mathbf{r}_0, \omega) & = \left.\frac{1}{2}\text{Re} \oint \textrm{d}\Omega\,\mathbf{\hat{n}} \cdot \mathbf{E}(\mathbf{r}, \omega) \times \mathbf{H}^*(\mathbf{r}, \omega)\right|_{\mathbf{r} \to \mathbf{r}_0}\nonumber\\
	& = \left|\frac{\mathbf{E}_2(\mathbf{r}_0, \omega)}{\mathbf{E}_0(\mathbf{r}_0, \omega)}\right|^2 \left[1 + \frac{6\pi c}{n \omega} \text{Im} \left(\mathbf{\hat{n}} \cdot \mathfrak{G}(\mathbf{r}_0, \mathbf{r}_0, \omega) \cdot \mathbf{\hat{n}}\right)\right],
\end{align}
where $\mathfrak{G}(\mathbf{r}_0, \mathbf{r}_0, \omega)$ is the Green's tensor, and the prefactor in the second line is the field enhancement factor, $A(\mathbf{r}_0, \omega) = \left|\mathbf{E}_2(\mathbf{r}_0, \omega) / \mathbf{E}_0(\mathbf{r}_0, \omega)\right|^2$. Similarly, the radiative power emitted in the presence of the sphere is
\begin{equation}
	\label{eq:05}
	P_\textrm{r}(\mathbf{r}_0, \omega) = \left.\frac{1}{2}\text{Re} \oint \textrm{d}\Omega\,\mathbf{\hat{n}} \cdot \mathbf{E}(\mathbf{r}, \omega) \times \mathbf{H}^*(\mathbf{r}, \omega)\right|_{r \to \infty}.
\end{equation}

These general expressions take, for a spherical NP, specific forms depending on the orientation of the transition dipole of the quantum system with respect to the NP. For a radially oriented transition dipole, the radiative and total emitted power become
\begin{widetext}
\begin{subequations}
\label{eqs:06}
\begin{equation}
\label{eq:06a}
	\frac{P^\perp_\text{r}(\mathbf{r}_0, \omega)}{P_0(\omega)} = A(\mathbf{r}_0, \omega) \frac{3}{2} \sum_n n(n + 1) (2 n + 1) \left|\frac{\psi_n(\rho_\text{s}) + a_n \zeta_n(\rho_\text{s})}{\rho_\text{s}^2}\right|^2,
\end{equation}
\begin{equation}
\label{eq:06b}
	\frac{P^\perp(\mathbf{r}_0, \omega)}{P_0(\omega)} = A(\mathbf{r}_0, \omega) \left[1 + \frac{3}{2} \sum_n n(n + 1) (2 n + 1) \text{Re} \left(\frac{a_n \zeta_n^2(\rho_\text{s})}{\rho_\text{s}^4}\right)\right],
\end{equation}
\begin{equation}
\label{eq:06c}
	\frac{P^\perp_\text{nr}(\mathbf{r}_0, \omega)}{P_0(\omega)} = \frac{P^\perp(\mathbf{r}_0, \omega)}{P_0(\omega)} - \frac{P^\perp_\text{r}(\mathbf{r}_0, \omega)}{P_0(\omega)},
\end{equation}
\end{subequations}
while the same quantities for a tangentially oriented dipole read
\begin{subequations}
\label{eqs:07}
\begin{equation}
\label{eq:07a}
	\frac{P^\parallel_\text{r}(\mathbf{r}_0, \omega)}{P_0(\omega)} = A(\mathbf{r}_0, \omega) \frac{3}{4} \sum_n (2 n + 1) \left[\left|\frac{\psi_n(\rho_\text{s}) + b_n \zeta_n(\rho_\text{s})}{\rho_\text{s}}\right|^2 + \left|\frac{\psi_n^\prime(\rho_\text{s}) + a_n \zeta_n^\prime(\rho_\text{s})}{\rho_\text{s}}\right|^2\right],
\end{equation}
\begin{equation}
\label{eq:07b}
	\frac{P^\parallel(\mathbf{r}_0, \omega)}{P_0(\omega)} = A(\mathbf{r}_0, \omega) \left[1 + \frac{3}{4} \sum_n (2 n + 1) \text{Re} \left(\frac{b_n \zeta_n^2(\rho_\text{s})}{\rho_\text{s}^2} + \frac{a_n \zeta_n^{\prime 2}(\rho_\text{s})}{\rho_\text{s}^2}\right)\right].
\end{equation}
\begin{equation}
\label{eq:07c}
	\frac{P^\parallel_\text{nr}(\mathbf{r}_0, \omega)}{P_0(\omega)} = \frac{P^\parallel(\mathbf{r}_0, \omega)}{P_0(\omega)} - \frac{P^\parallel_\text{r}(\mathbf{r}_0, \omega)}{P_0(\omega)},
\end{equation}
\end{subequations}
\end{widetext}

In all the above expressions the summation is over the multipole moments $n$ of the sphere, $\psi_n$ and $\zeta_n$ represent the first and third kind Ricatti-Bessel functions, $\rho_\text{s} = k r_0$, and $a_n$ and $b_n$ are the Mie scattering coefficients of the sphere. The relative decay rates can now be calculated from Eq.~\eqref{eq:01}.

\subsection{Energy Transfer Rate}
\label{sub:II.B}

The energy transfer rate between donor and acceptor quantum systems can be calculated similar to the decay rates from the expression
\begin{equation}
\label{eq:08}
	\frac{\gamma_\text{DA}(\mathbf{r}_\text{A}, \mathbf{r}_\text{D}, \omega)}{\gamma_0(\omega)} = \frac{P_\text{DA}(\omega)}{P_0(\omega)}
\end{equation}
where $P_\text{DA}(\omega)$ represents the power emitted by the donor D and absorbed by the acceptor A.

It can be shown that the energy transfer (ET) rate between a donor-acceptor pair modeled as point dipoles, is given as\cite{Zhang2014}
\begin{widetext}
\begin{align}
\label{eq:09}
	\gamma_{\text{DA}}(\mathbf{r}_\text{A}, \mathbf{r}_\text{D}) & = \int\limits_0^\infty \text{d}\omega f_\text{D}(\omega) \gamma_\text{DA}(\mathbf{r}_\text{A}, \mathbf{r}_\text{D}, \omega) \nonumber \\
	& = 18 \pi \frac{Y_D}{\tau_\text{D}} \int\limits_0^{\infty} \text{d}\omega f_\text{D}(\omega) |\mathbf{n}_\text{A} \cdot  \mathfrak{G}(\mathbf{r}_{\text{A}}, \mathbf{r}_{\text{D}}, \omega) \cdot \mathbf{n}_{\text{D}}|^2 \sigma_{\text{A}}(\omega).
\end{align}
\end{widetext}
where $\tau_\text{D}$ is the donor lifetime in the absence of the acceptor, $Y_\text{D}$ is the quantum yield of the donor, $f_\text{D}(\omega)$ is the area-normalized donor emission spectrum, i.e.~$\int_0^\infty\text{d}\omega f_\text{D}(\omega) = 1$, $\sigma_\text{A}(\omega)$ is the acceptor absorption cross-section and $\mathfrak{G}(\mathbf{r}_\text{A}, \mathbf{r}_\text{B}, \omega)$ is the Green's tensor in the particular geometry. Finally, $\mathbf{n}_\text{D(A)}$ is a unit vector along the direction of the transition dipole moment of the donor (acceptor).

\section{Results and Discussion}
\label{sec:III}

\subsection{Decay Rates}
\label{sub:III.A}

A question of some interest and which has also generated some discussion in the literature regards the precise nature of the energy transfer process between a donor quantum system (fluorescent dye, quantum dot) and a small Au NP. One candidate mechanism is FRET,\cite{Forster1948} which assumes that the Au NP can be regarded as a polarizable dipole and the energy transfer process is, therefore, that between two point dipoles, with a characteristic $r^{-6}$ dependence on the donor-acceptor distance, $r$. A second mechanism is NSET,\cite{Singh2007,Singh2010} which views the energy transfer process as occuring between a donor dipole and an infinite planar surface, with a characteristic $d^{-4}$ dependence on the donor-surface distance, $d$.\cite{Persson1981,Persson1982}

Recently, Sukharev \textit{et al.}\cite{Sukharev2014} have calculated the distance dependence of the non-radiative decay rates of point dipoles near Ag nanospheres with a diameter of 40 nm. The results of their calculations have shown various distance dependence regimes, accommodating both FRET and NSET type dependences. In this section we consider the distance, as well as emission wavelength, quantum yield, and Au NP size dependence of the energy transfer rate between an emitter and Au nanospheres of sizes ranging from subnanometer, when no LSP is supported, to a few hundred nanometers, well in the radiative regime.

\subsubsection{Experimental Verification}
\label{subsub:III.A.1}

\begin{figure*}[t]
\centering
	\subfloat[Schematic\label{fig:01a}]{\includegraphics[height=0.3\textwidth]{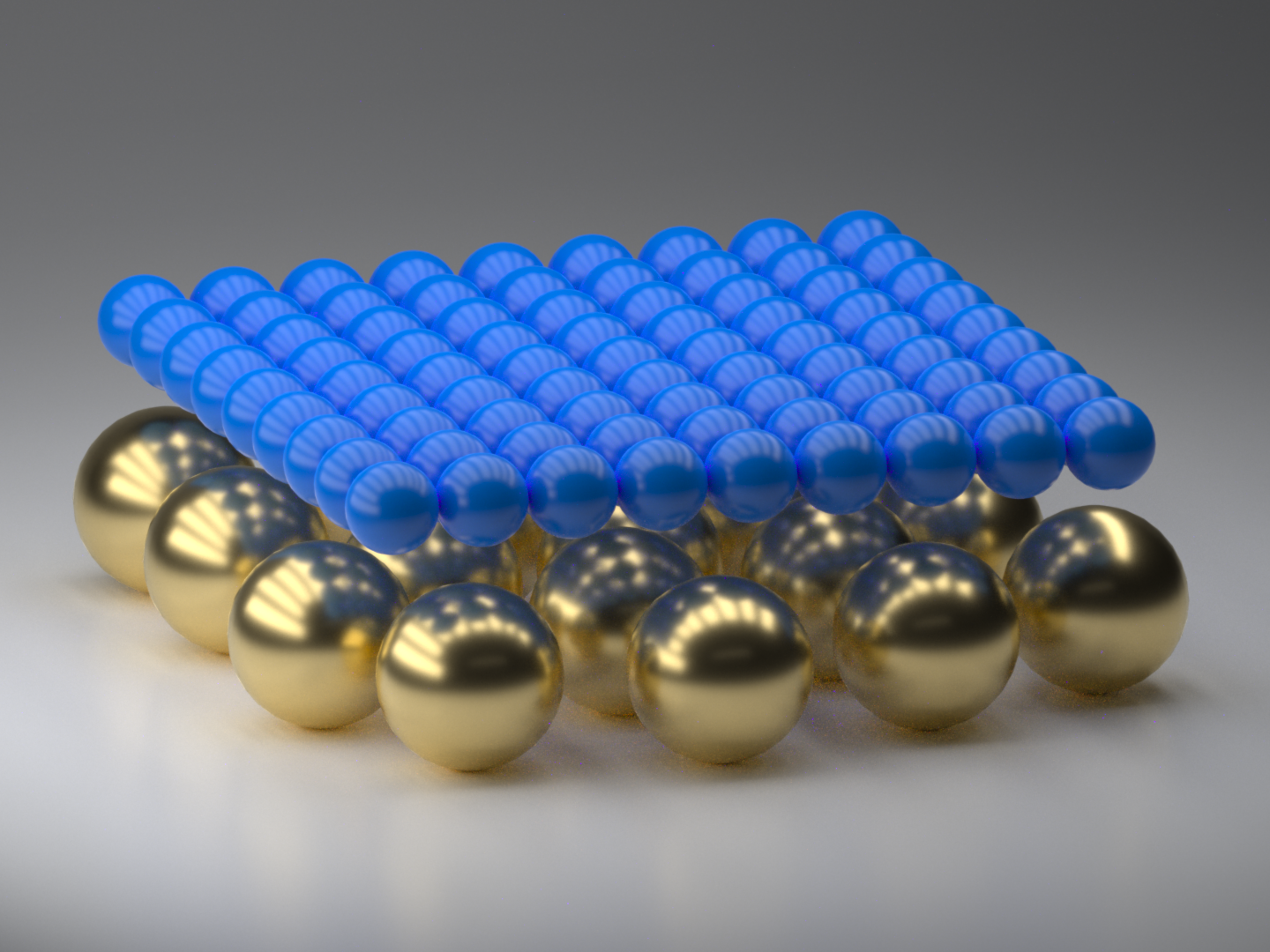}}\hspace{5mm}
	\subfloat[Optical properties\label{fig:01b}]{\includegraphics[height=0.3\textwidth]{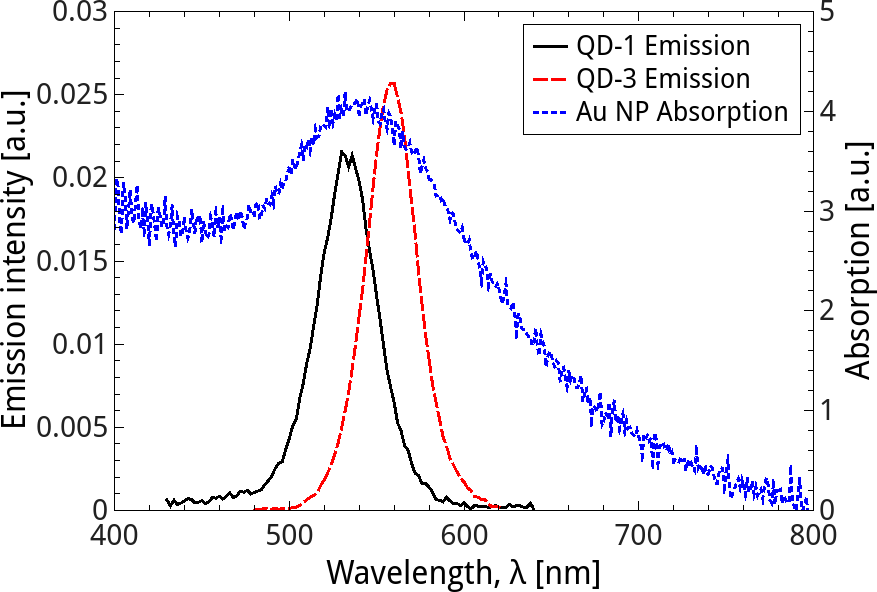}}\\
	\subfloat[QD-1\label{fig:01c}]{\includegraphics[height=0.3\textwidth]{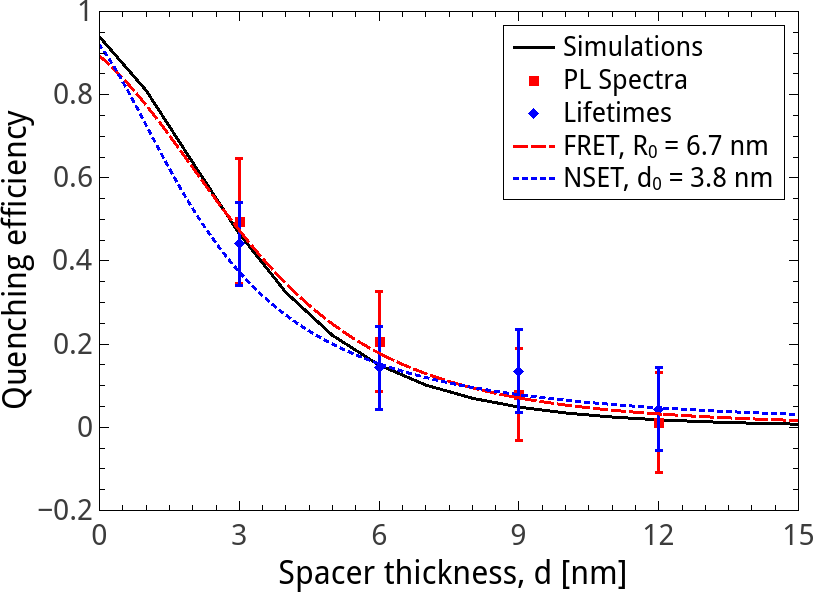}}\hspace{5mm}
	\subfloat[QD-3\label{fig:01d}]{\includegraphics[height=0.3\textwidth]{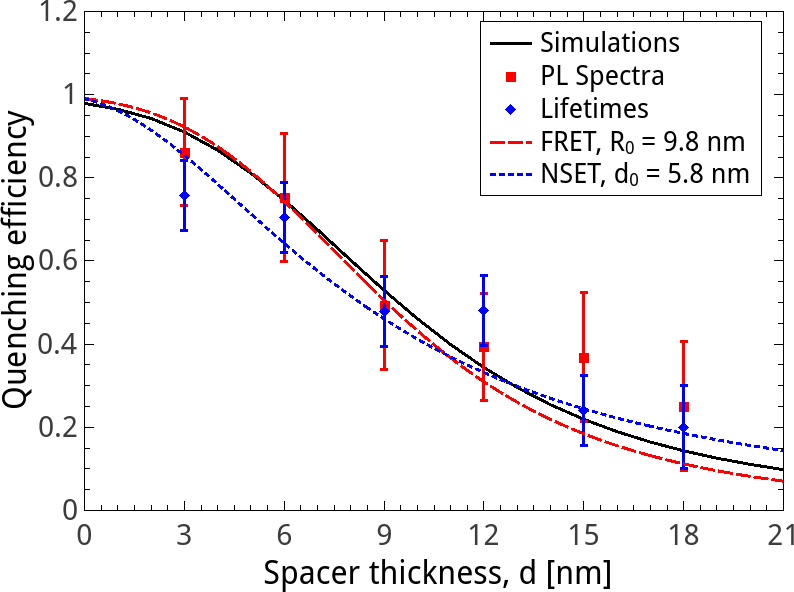}}\\
	\caption{(Color online) Experimental verification of the model used in this paper; (a) schematic of the QD monolayer (blue spheres) on top of the Au NP monolayer (gold spheres); (b) normalized emission spectra of the two quantum dot samples, as well as the absorption spectrum of the Au NPs. The experimental data are taken from Ref.~\onlinecite{Zhang2012} for the quantum dots labeled as (c) QD-1 and (d) QD-3 in that reference. The black curve shows our parameter-free simulation. The red dashed curve is a fit of the experimental data with a FRET model, while the blue dotted curve corresponds to a fit of the experimental data with a NSET model.\label{fig:01}}
\end{figure*}

We first verify our model by considering experimental measurements performed in our lab and reported in Ref.~\onlinecite{Zhang2012}. Panel~\ref{fig:01a} of Fig.~\ref{fig:01} shows a schematic of the experimental samples, prepared using a Layer-by-Layer technique.
A monolayer of Au nanospheres with a radius $a = 2.75$ nm is separated by a polyelectroyte layer of thickness $d$ (not shown here) from a monolayer of closely packed quantum dots (blue spheres). Panel~\ref{fig:01b} presents the optical properties of the quantum dots and the Au NPs. The black and red curves represent the emission spectra of two types of quantum dots, labeled QD-1 and QD-3 in Ref.~\onlinecite{Zhang2012} and in the rest of this paper, whereas the blue curve represents the extinction spectrum of the Au NP monolayer. The photoluminescence (PL) spectra and the time-resolved PL decays of these samples were recorded. The quenching of the QD emission was calculated from these data as follows. The PL quenching was calculated as $Q_\text{PL} = 1 - I_\text{onAu} / I_\text{QD}$, where $I_\text{onAu}$ ($I_\text{QD}$) represents the integrated spectral emission of the QD monolayer in the presence (absence) of the Au NP monolayer. Similarly, the lifetime quenching was calculated as $Q_\text{LT} = 1 - \tau_\text{onAu} / \tau_\text{QD}$, with $\tau_\text{onAu}$ ($\tau_\text{QD}$) being the PL decay times of the QD monolayer in the presence (absence) of the Au NP monolayer. Panels~\ref{fig:01c} and \ref{fig:01d} of Fig.~\ref{fig:01} show the PL and lifetime quenching efficiencies as red squares and blue diamonds, respectively, for the two types of quantum dots, QD-1 and QD-3. The agreement between these two sets of experimental data, within the error bars, shows that the radiative decay rate of the quantum dots is not modified significantly by the presence of the Au NP monolayer. We also show two theoretical fits of the experimental data with the FRET model (red dashed curve) and the NSET model (blue dotted curve).

The results of our simulation are shown as a continuous black line in panels \ref{fig:01c} and \ref{fig:01d}. The simulations have been performed using the experimental values of the quantum dot properties -- size, quantum yield, emission spectrum -- and Au NP properties -- size and concentration in the monolayer. As such, the simulation uses no free parameters and one can appreciate the close agreement between simulation and experiment. The theoretical quenching efficiency is calculated as
\begin{equation}
\label{eq:10}
	Q_\text{th} = \frac{\gamma_\text{nr}}{\gamma_\text{r} + \gamma_\text{nr}^0 + \gamma_\text{nr}},
\end{equation}
where $\gamma_\text{r}$ and $\gamma_\text{nr}$ are the radiative and non-radiative decay rates of the quantum dots in the presence of the Au NP monolayer, while $\gamma_\text{nr}^0$ is the intrinsic non-radiative decay rate of the quantum dots, when the quantum yield is less than 100\%. In this interpretation, $\gamma_\text{nr}$ is the energy transfer rate from the donor quantum dot to the acceptor Au NPs. While both theoretical fits and our simulations agree with the experimental data within the error bars, the characteristic distances extracted from the FRET fit do not agree with those calculated from the spectral overlap (see \onlinecite{Zhang2012} for a more in-depth discussion). There are many experiments suggesting that the energy transfer mechanism from the quantum dot to the Au NP has a more complex distance dependence than these models allow for. We will investigate this distance dependence in section \eqref{subsub:III.A.3}.

\subsubsection{Emission Wavelength and Quantum Yield Dependence of the Decay Rates}
\label{subsub:III.A.2}

Now that the theoretical model and simulation procedure have been validated by experimental measurements, we consider the simple case of a single emitter placed close to a single Au NP. We investigate the influence of the emission wavelength of the emitter and its intrinsic quantum yield on the quenching efficiency due to energy transfer to the Au NP. The emission spectrum of the quantum emitter will be modeled as a gaussian distribution with a varying central wavelength and a full-width at half-maximum (FWHM) of $\approx 45$ nm, comparable to the experimental emission spectra from Fig.~\ref{fig:01c}.

\begin{figure*}[t]
	\subfloat[$a = 0.25$ nm\label{fig:02a}]{\includegraphics[height=0.22\textwidth]{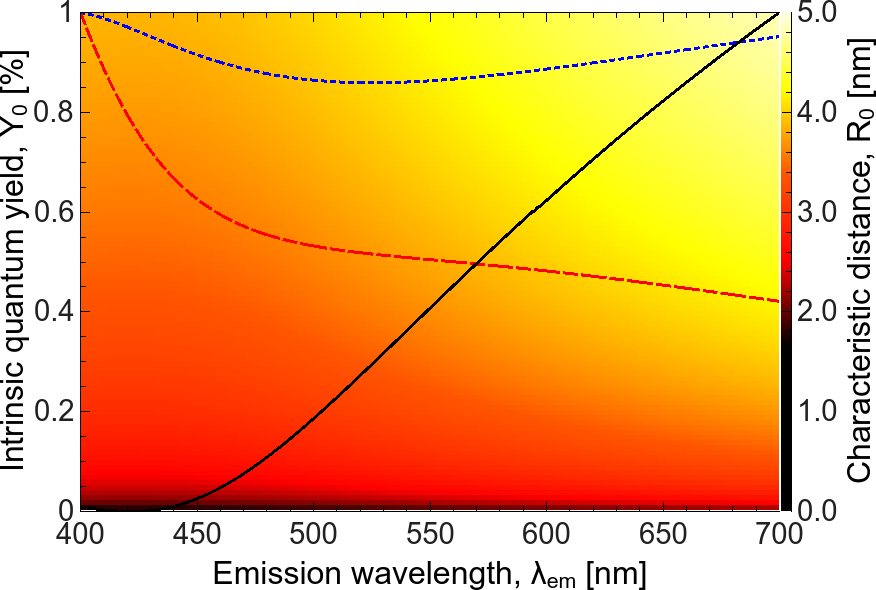}}\hfill
	\subfloat[$a = 2.5$ nm\label{fig:02b}]{\includegraphics[height=0.22\textwidth]{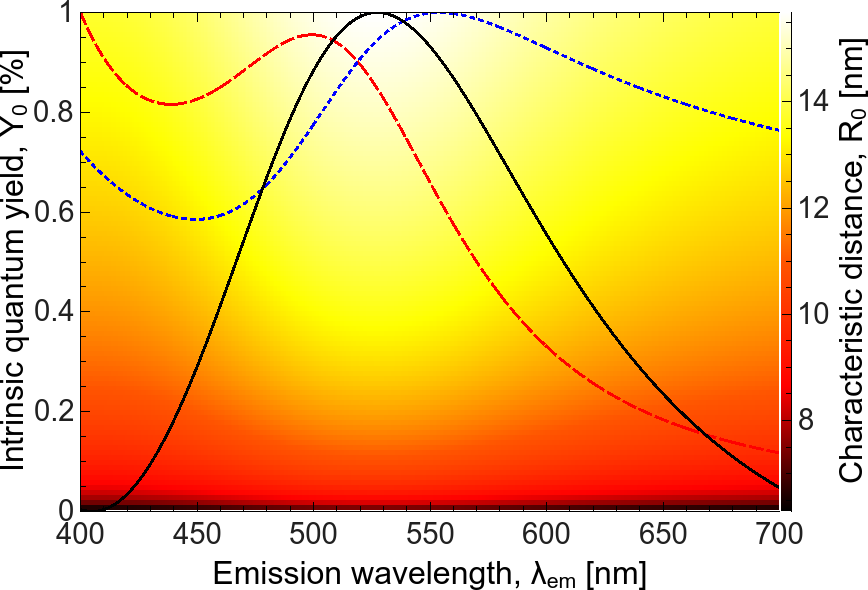}}\hfill
	\subfloat[$a = 5.0$ nm\label{fig:02c}]{\includegraphics[height=0.22\textwidth]{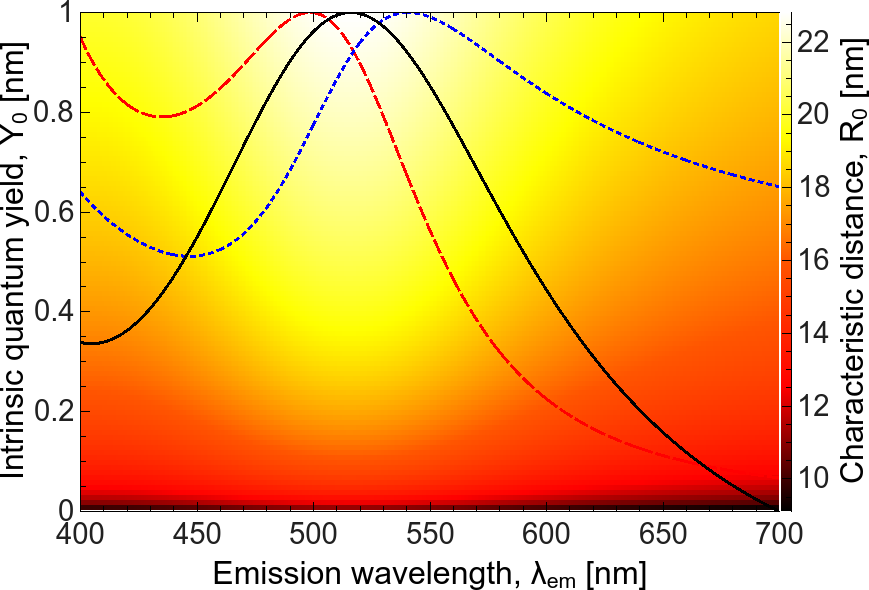}}\\
	\subfloat[$a = 7.5$ nm\label{fig:02d}]{\includegraphics[height=0.22\textwidth]{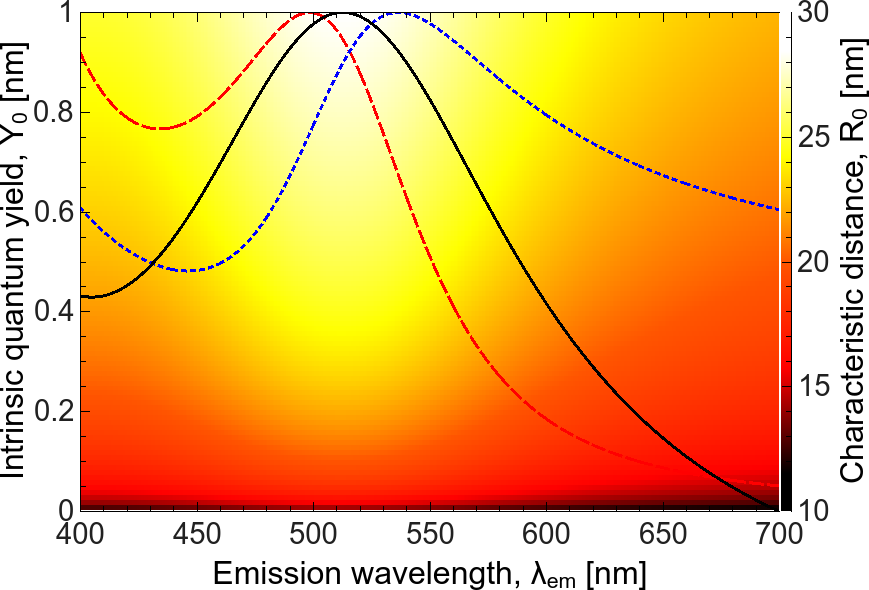}}\hfill
	\subfloat[$a = 10$ nm\label{fig:02e}]{\includegraphics[height=0.22\textwidth]{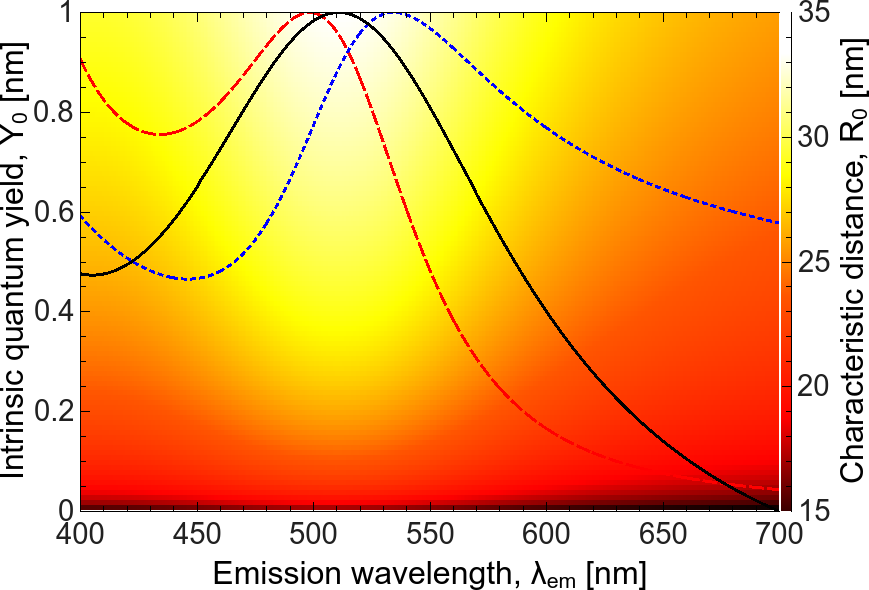}}\hfill
	\subfloat[$a = 20$ nm\label{fig:02f}]{\includegraphics[height=0.22\textwidth]{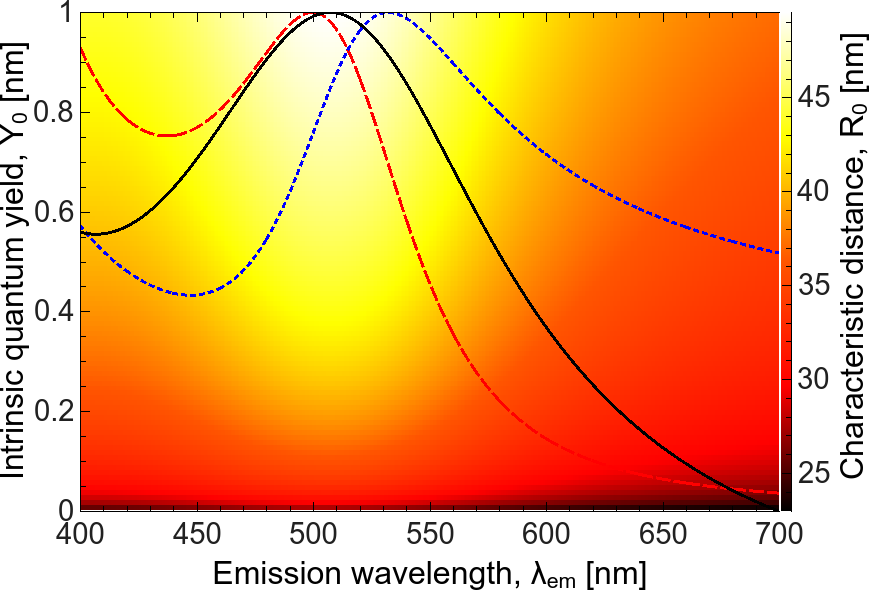}}\\
	\caption{(Color online) Emission wavelength and intrinsic quantum yield dependence of the characteristic distance, $R_0$, of a quantum emitter near Au NP of several radii, from (a) $a = 0.25$ nm to (f) $a = 20$ nm. The excitation wavelength is $\lambda_{\textrm{exc}} = 400$ nm. The solid black line corresponds to a fixed value of the quantum yield, $Y_0 = 100\%$, the dashed red line is the extinction efficiency and the blue dotted line is the near-field scattering efficiency for each Au NP.\label{fig:02}}
\end{figure*}

The quenching efficiency of the quantum emitter close to a single Au NP can depend strongly on the emission wavelength of the quantum emitter and its intrinsic quantum yield. Fig.~\ref{fig:02} shows the characteristic 50\% quenching distance, $R_0$, as a function of the emission wavelength of the quantum emitter, $\lambda_{\textrm{em}}$, and its intrinsic quantum yield, $Y_0$, for several sphere sizes. For the smallest Au NP size, $a = 0.25\;\text{nm}$ (Fig.~\ref{fig:02a}), the dependence of $R_0$ on $\lambda_{\text{em}}$ and $Y_0$ is monotonic, since a Au NP of this size does not support a LSP.

When considering a sphere of radius $a = 2.5\;\text{nm}$ (Fig.~\ref{fig:02b}), the characteristic distance, $R_0$, attains a maximum value at a wavelength $\lambda_\text{em} = 527\;\text{nm}$, between the peaks of the extinction and the near-field scattering efficiency spectra. The spectral dependence of $R_0$, for a 100\% quantum yield is illustrated by the solid black line. Alongside this quantity, we have also included the extinction efficiency (red dashed curve) and the near-field scattering efficiency (blue dotted curve). The values of these efficiencies are unimportant for the present discussion, and they are shown as normalized quantities.

As the sphere size is increased from $a = 2.5\;\text{nm}$ (Fig.~\ref{fig:02b}) to $a = 5.0\;\text{nm}$ (Fig.~\ref{fig:02c}), $a = 7.5\;\text{nm}$ (Fig.~\ref{fig:02d}), $a = 10\;\text{nm}$ (Fig.~\ref{fig:02e}) and $a = 20.0\;\text{nm}$ (Fig.~\ref{fig:02f}), the dependence of $R_0$ on $\lambda_{\text{em}}$ undergoes a blue-shift from $\lambda_\text{em} = 527\;\text{nm}$ ($a = 2.5\;\text{nm}$) to $\lambda_\text{em} = 516\;\text{nm}$ ($a = 5.0\;\text{nm}$), $\lambda_\text{em} = 513\;\text{nm}$ ($a = 7.5\;\text{nm}$), $\lambda_\text{em} = 511\;\text{nm}$ ($a = 10.0\;\text{nm}$) and $\lambda_\text{em} = 507\;\text{nm}$ ($a = 20.0\;\text{nm}$).

The blue-shift of the maximum of the characteristic distance, $R_0$, with sphere size is particularly visible when comparing panels \ref{fig:02c} and \ref{fig:02f}. For the smaller spheres, the maximum in $R_0$ lies between the maximum of the extinction and the near-field scattering efficiencies (\ref{fig:02c}). As the sphere size is increased, the near-field scattering becomes less important, while extinction (absorption plus far-field scattering) begins to dominate, and the maximum of $R_0$ moves towards the peak of the extinction efficiency (\ref{fig:02f}).

There are a number of points to be made regarding the joint dependence of the characteristic distance, $R_0$, on the emission wavelength, $\lambda_{\textrm{em}}$, and the intrinsic quantum yield, $Y_0$. Whereas $R_0$ increases monotonically with the intrinsic quantum yield of the emitter, as expected, its dependence on the emission wavelength exhibits a maximum in the vicinity of the LSP peak. It is, therefore, important, that in experiments investigating the emission wavelength dependence of the quenching efficiency of a quantum emitter, one ascertain whether the intrinsic quantum yield of the quantum emitter also changes, as changes in this variable may determine the wavelength of the maximum in $R_0$ that one observes.

\subsubsection{Distance Dependence of the Decay Rates}
\label{subsub:III.A.3}

To investigate the dependence of the non-radiative decay rates on the distance between the quantum emitter and the center of the Au NP, $r$, we write the non-radiative decay rate as:
\begin{equation}
\label{eq:11}
	\frac{\gamma_\text{nr}(\mathbf{r})}{\gamma_0} = \left(\frac{R_0}{r}\right)^n,
\end{equation}
where $R_0$ is a constant to be determined -- the so-called characteristic distance -- and $n$ is an exponent, taking the value $n = 6$ for the FRET model. Taking the logarithm of both sides of Eq.~\eqref{eq:11} above, one can rewrite it as
\begin{equation}
\label{eq:12}
	\log\left[\frac{\gamma_\text{nr}(\mathbf{r})}{\gamma_0}\right] = n\log R_0 - n\log r.
\end{equation}
The first derivative with respect to $\log r$ of the right-hand side of the above equation will give $n$ as a function of the distance from the quantum emitter to the center of the Au NP, $n = n(r)$.

The emission wavelength of the quantum emitter has been fixed at $\lambda_\text{em} = 525$ nm, with an intrinsic quantum yield of $Y_0 = 100\%$.

\begin{figure*}[t]
	\subfloat[a = 0.25 nm\label{fig:03a}]{\includegraphics[height=0.225\textwidth]{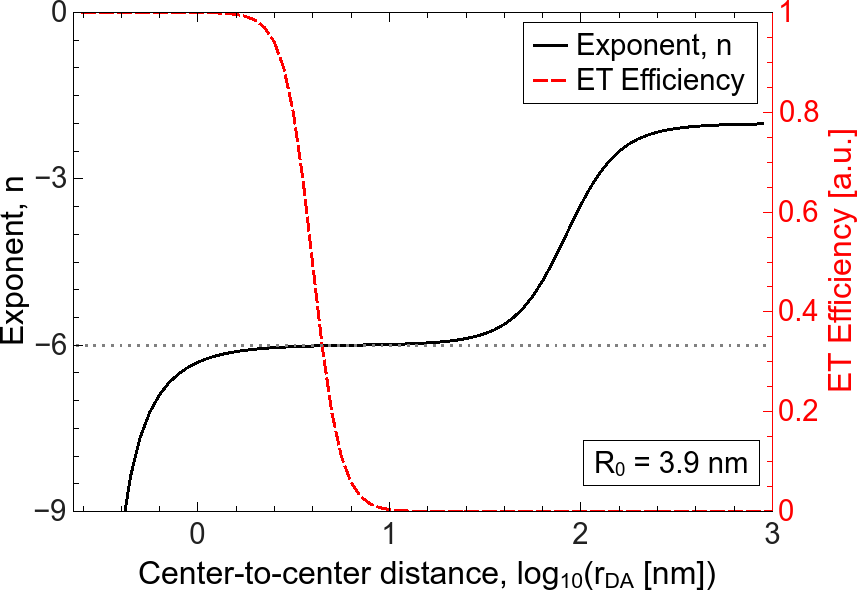}}\hfill
	\subfloat[a = 2.5 nm\label{fig:03b}]{\includegraphics[height=0.225\textwidth]{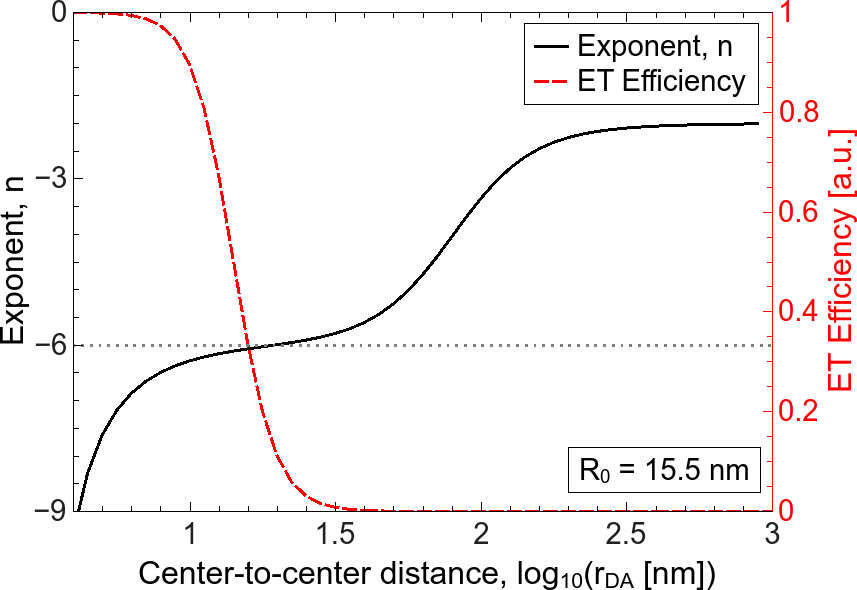}}\hfill
	\subfloat[a = 5.0 nm\label{fig:03c}]{\includegraphics[height=0.225\textwidth]{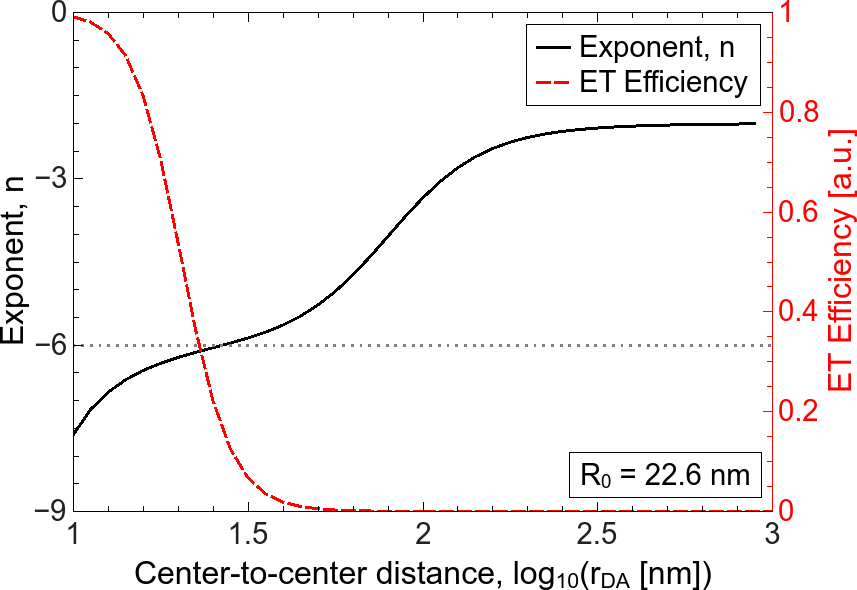}}\\
	\subfloat[a = 7.5 nm\label{fig:03d}]{\includegraphics[height=0.225\textwidth]{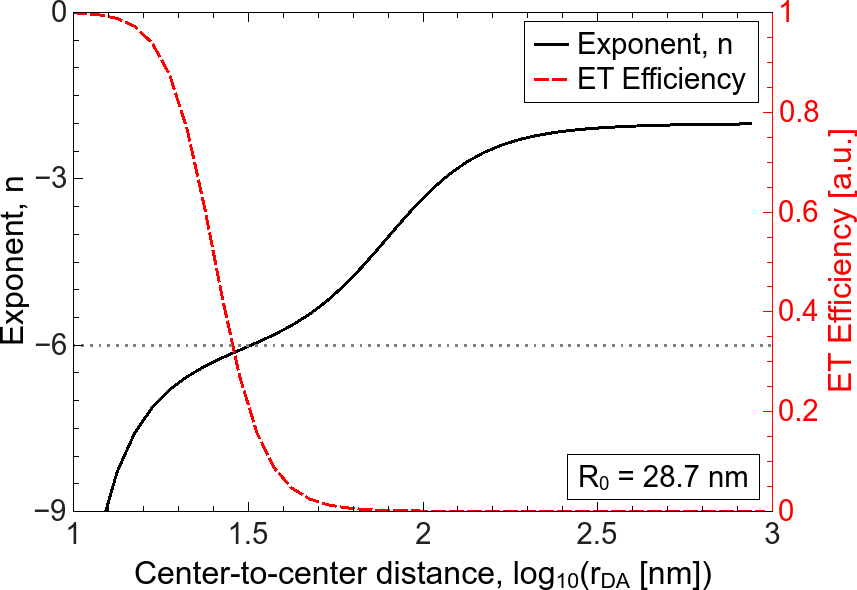}}\hfill
	\subfloat[a = 10.0 nm\label{fig:03e}]{\includegraphics[height=0.225\textwidth]{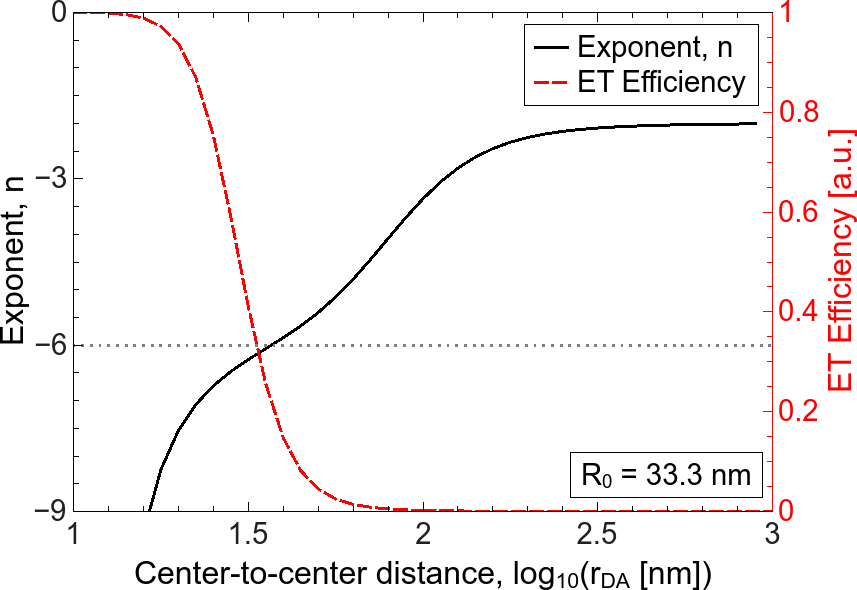}}\hfill
	\subfloat[a = 20.0 nm\label{fig:03f}]{\includegraphics[height=0.225\textwidth]{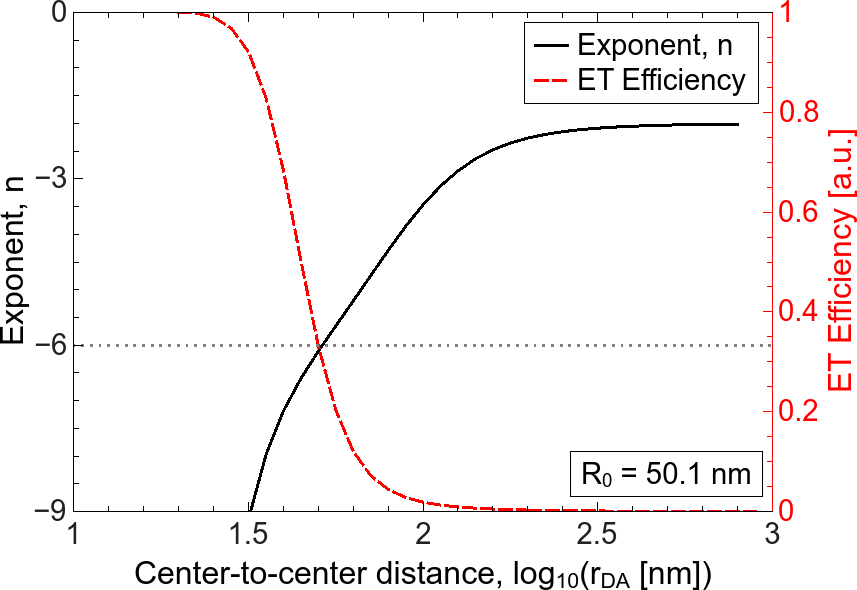}}\\
	\caption{(Color online) Exponent $n$ and the non-radiative decay rate efficiency as a function of distance for Au NP with radii (a) 0.25 nm, (b) 2.5 nm, (c) 5 nm, (d) 7.5 nm, (e) 10 nm and (f) 20 nm.\label{fig:03}}
\end{figure*}

We have considered the distance dependence of the non-radiative decay rate of a quantum emitter in the presence of Au nanospheres with small radii, between 0.25 and 20 nm. Fig.~\ref{fig:03} shows the exponent $n$ of the distance dependence of the non-radiative decay rate of the quantum emitter as a function of the distance of the emitter to the center of the Au NP.
Also shown is the energy transfer efficiency (dashed lines) in the panels of Fig.~\ref{fig:03}. As this figure shows, the non-radiative decay rate follows a $r^{-6}$ distance dependence over a range of distances for which the ET efficiency is around 50\%. This is most clear for the smaller NP, with a radius $a = 0.25$ nm (Fig.~\ref{fig:03a}), but it is visible for the larger ones as well (Fig.~\ref{fig:03b}), over a smaller range. Given that the $x$-axis is logarithmic, this range is still quite large. As the radius of the Au NP is increased, however, the $r^{-6}$ distance dependence regime becomes less dominant until it completely disappears for panel \ref{fig:03f} in the figure, for which the radius of the Au NP is 20 nm. One can, therefore, infer a FRET-type mechanism for the energy transfer rate from the emitter to the Au NP, when the distance of the emitter to the surface of the Au NP is larger than the diameter of the NP. The energy transfer rate deviates from a simple dipole pair FRET model when the emitter-Au NP surface separation is smaller than the diameter of the Au NP, in which case the higher multipole orders in the electromagnetic response of the NP have a non-negligible contribution to the ET rate and the NP can no longer be accurately approximated as a point dipole. The characteristic distances, $R_0$, shown in the legends of Fig.~\ref{fig:03} are the 50\% quenching efficiency distances. The value of the quantum yield will not affect the exponent, but it will decrease the $R_0$ distance, according to $R_0^\prime = R_0\sqrt[6]{Y_\text{D}^0}$.

\subsubsection{Au NP Size Dependence of the Decay Rates}
\label{subsub:III.A.4}

\begin{figure*}[t]
	\subfloat[$Y_0 = 100\%$\label{fig:04a}]{\includegraphics[height=0.23\textwidth]{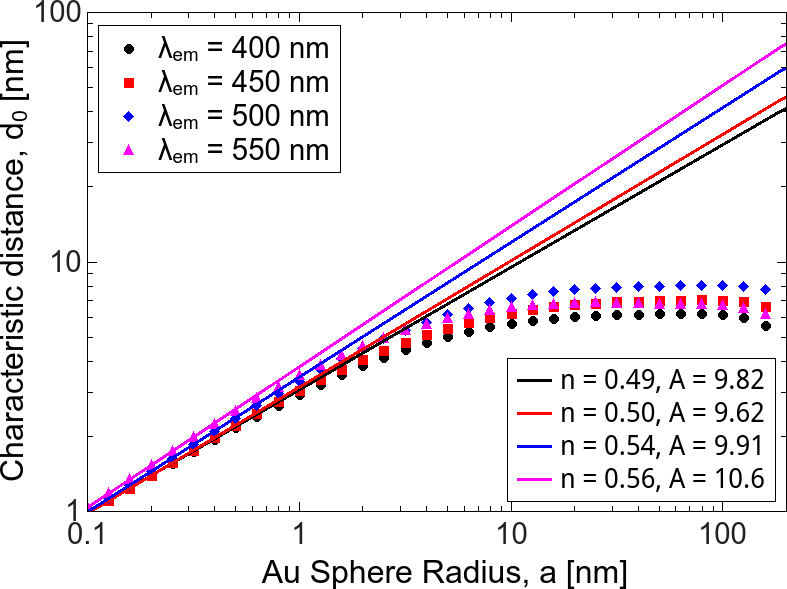}}\hfill
	\subfloat[$Y_0 = 10\%$\label{fig:04b}]{\includegraphics[height=0.23\textwidth]{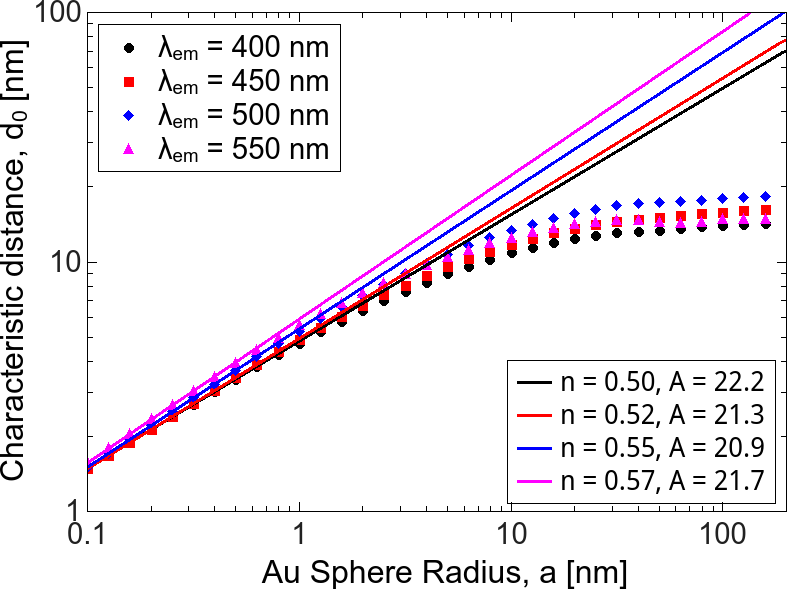}}\hfill
	\subfloat[$Y_0 = 1\%$\label{fig:04c}]{\includegraphics[height=0.23\textwidth]{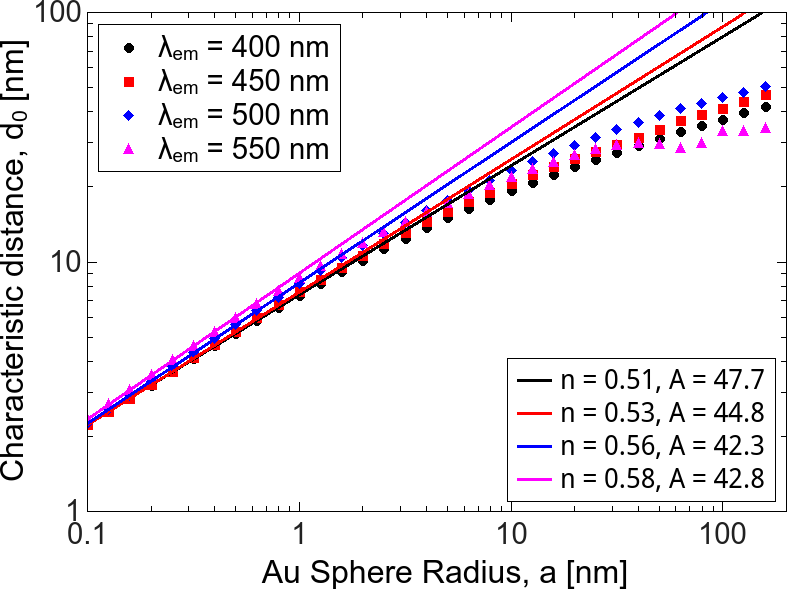}}
	\caption{(Color online) Dependence of the characteristic distance, $R_0$, on the Au sphere radius, $a$, for three different emitter intrinsic quantum yields, (a) $Y_0 = 100\%$, (b) $Y_0 = 10\%$ and (c) $Y_0 = 1\%$, and several emission wavelengths of the emitter. Both axes are logarithmic.\label{fig:04}}
\end{figure*}

Finally, we now consider the dependence of the characteristic distance on the size of the Au NP. For this investigation, however, we will not consider the distance of the quantum emitter to the center of the Au NP, $R_0$, but to its surface, denoted $d_0$. The reason for this change is that, as the size of the Au NP is increased, $R_0$ is no longer a good measure of the 50\% quenching distance, since it is almost equal to the Au NP radius, $a$. Fig.~\ref{fig:04} shows the dependence of $d_0$ on $a$ for three different intrinsic quantum yields of the quantum emitter: (\ref{fig:04a}) $Y_0 = 100\%$, (\ref{fig:04b}) $Y_0 = 10\%$ and (\ref{fig:04c}) $Y_0 = 1\%$.
Additionally for each panel, we have also calculated this dependence for several emission wavelengths of the quantum emitter, $\lambda_\text{em} =$ 400 nm, 450 nm, 500 nm, and 550 nm, spanning the LSP range. The log-log scale in Fig.~\ref{fig:04} shows that the dependence of $d_0$ on $a$ follows a power law of the form
\begin{equation}
\label{eq:13}
	d_0 = \left(A a\right)^n 
\end{equation}
with two regimes. For the first regime, valid at small Au NP radii, the exponent $n$ is in the range 0.5--0.6, while $A$ is in the range 10--50. For the second regime, valid at much larger Au NP radii, the characteristic distance to the surface reaches a saturation regime, where it becomes independent of the radius, $a$. The exponent is then $n = 0$ and we have $A \approx 10$ nm (\ref{fig:04a}), much smaller than $a$. In this latter regime quenching of the quantum system emission by the Au NP is significant only extremely close to the Au NP surface. As soon as the quantum system is removed from the surface of the Au NP, the emission quenching will no longer be of any importance, as the emission properties of the quantum system will be dominated by scattering from the Au NP.

For the first regime, of small Au NP radii, the dependence of $d_0$ on $a$ is approximately of the form $d_0 = \sqrt{A a}$. As can be seen from the figure, the dependence of $d_0$ on the emission wavelength is rather weak. The dependence on the quantum yield, $Y_0$ follows the $\sqrt[6]{Y_0}$ dependence mentioned in the previous subsection.

\subsection{Energy Transfer Rates}
\label{sub:III.B}

Next we consider the energy transfer rates between a donor-acceptor pair placed in close proximity to a Au nanosphere.

\subsubsection{Experimental Verification}
\label{subsub:III.B.1}

\begin{figure*}[t]
	\subfloat[Schematic\label{fig:05a}]{\includegraphics[height=0.22\textwidth]{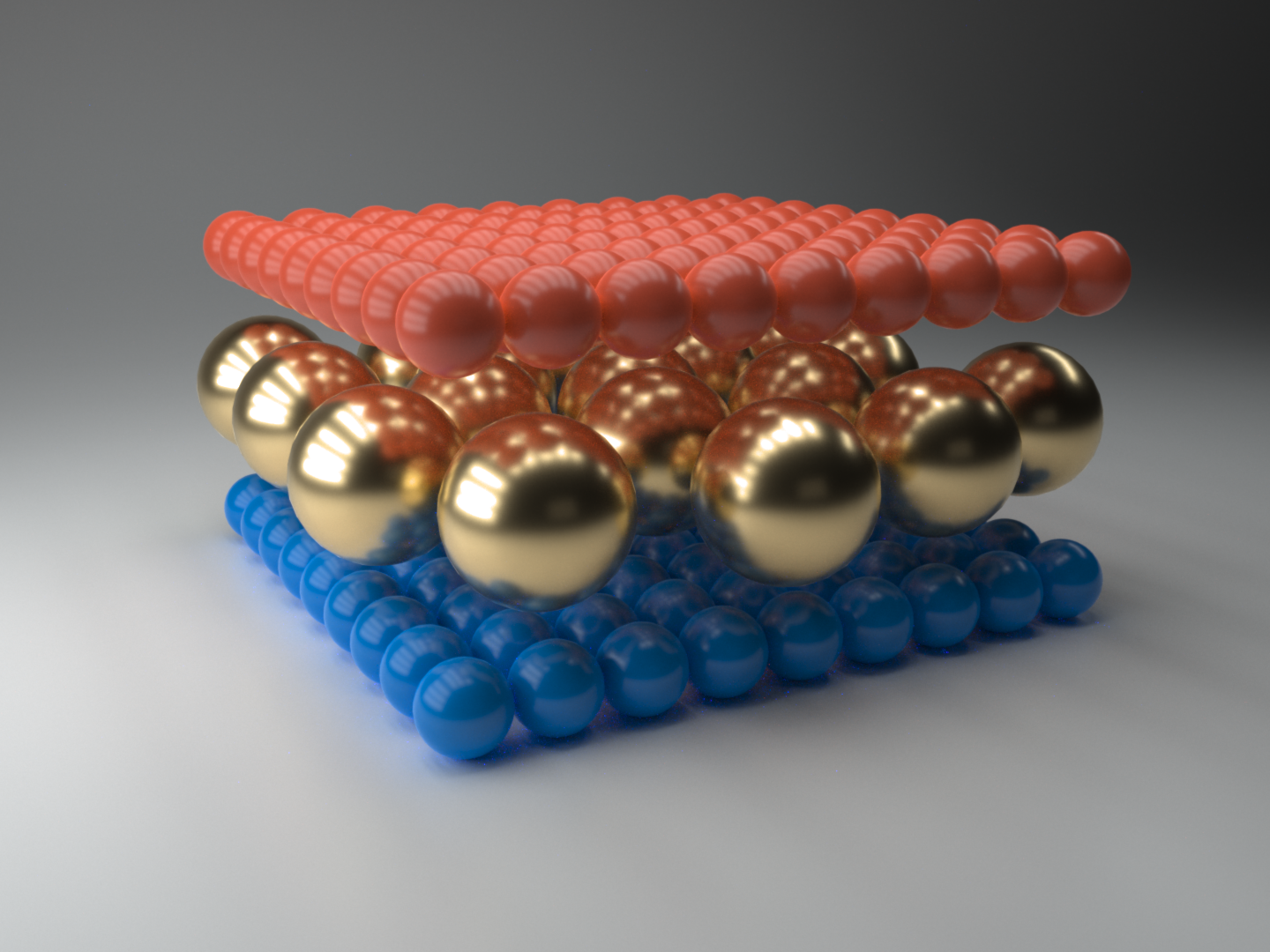}}\hspace{4mm}
	\subfloat[Optical properties\label{fig:05c}]{\includegraphics[height=0.22\textwidth]{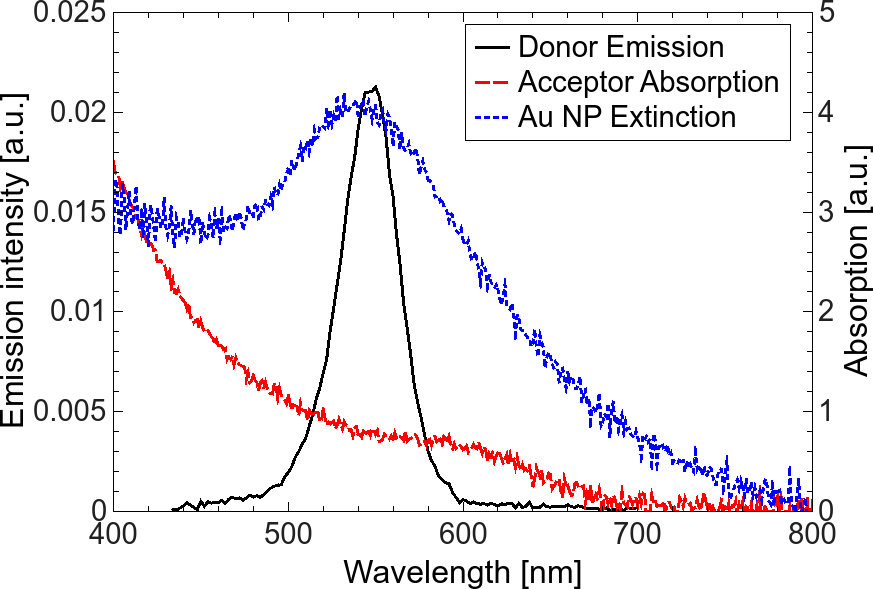}}\hspace{4mm}
	\subfloat[Experimental results\label{fig:05b}]{\includegraphics[height=0.22\textwidth]{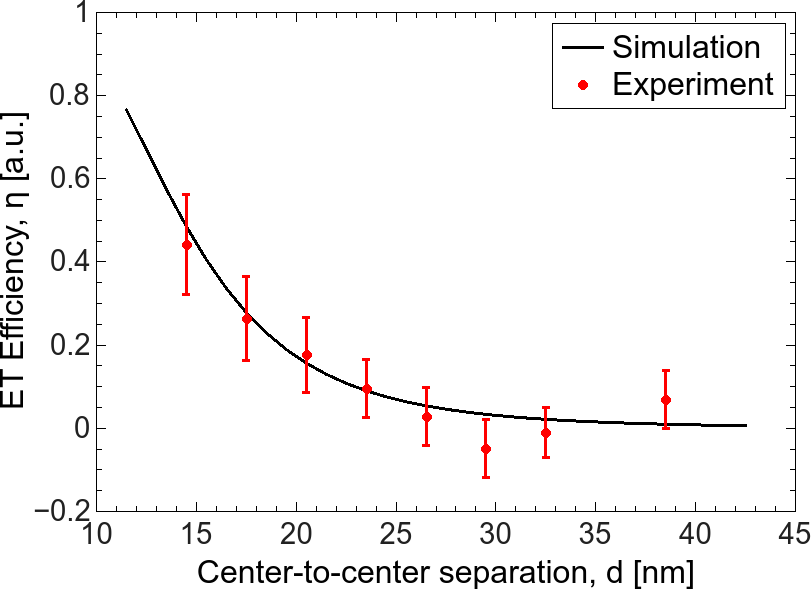}}\\
	\caption{(Color online) (a) Schematic of the Au nanosphere monolayer (radius 2.75 nm) sandwiched between the donor QD monolayer (blue spheres) and the acceptor QD monolayer (red spheres); (b) Emission spectrum of donor QD monolayer (black solid line), absorption spectrum of the acceptor QD monolayer (red dashed line) and extinction of Au NP monolayer (blue dotted line) (c) Energy transfer efficiency \textit{versus} the center-to-center distance for monolayers of quantum dots separated by a monolayer of Au nanospheres of radius 2.75 nm. The donor-Au NP separation is kept fixed at 3 nm. See reference \onlinecite{Zhang2014}. The black solid line represents our simulation results.\label{fig:05}}
\end{figure*}

We validated the model by comparing it with experimental measurements performed in our laboratory on energy transfer rates between monolayers of donor and acceptor quantum dots, separated by a monolayer of Au nanospheres.\cite{Zhang2014} The schematic of the structure is presented in Fig.~\ref{fig:05a}, where the donor quantum dots are represented as blue spheres in the bottom monolayer, the acceptor quantum dots are represented as red spheres in the top monolayer, and the Au NPs are represented as gold spheres in the middle monolayer. The separation of the different monolayers is controlled by the use of polyelectrolyte layers of known thickness (not shown in the schematic). Fig.~\ref{fig:05c} shows the emission spectrum of the donor quantum dot monolayer (black continuous line), the absorption spectrum of the acceptor quantum dot monolayer (red dashed line) and the extinction spectrum of the Au NP monolayer, showing the LSP peak (blue dotted line). This panel shows a good overlap between the different spectra, an essential feature for optimizing the energy transfer process. Finally, fig.~\ref{fig:05b} shows the calculated energy transfer efficiency in such a system, together with the experimental measurements.
In this multilayer structure, the donor-Au NP separation is kept fixed at 3 nm, and the acceptor-Au NP separation is varied. The simulation uses experimental values of all parameters and is, therefore, parameter-free. The experimental results are shown as red discs, together with the associated error bars, while the results of the simulations are presented as the black solid line. A good agreement between the model and experiment is observed. We feel justified, therefore, to now use this model in the following investigation of the distance, angular position, spectral overlap and Au NP size dependencies of the energy transfer in the presence of Au nanospheres.

\subsubsection{Distance Dependence of the Energy Transfer Efficiencies}
\label{subsub:III.B.2}

Starting with this section, we consider a single donor-Au NP-acceptor triad and investigate the donor-acceptor energy transfer. In this, as well as the following sections, the intrinsic quantum yield of the donor is, $Y_0^D = 100\%$, the donor emission spectrum is modelled as a gaussian with a FWHM $\approx 45$ nm, while we use the experimental absorption spectrum of the acceptor shown in panel \ref{fig:05c}.

\begin{figure*}[t]
	\subfloat[$a = 2.5$ nm, fixed donor distance\label{fig:06a}]{\includegraphics[height=0.26\textwidth]{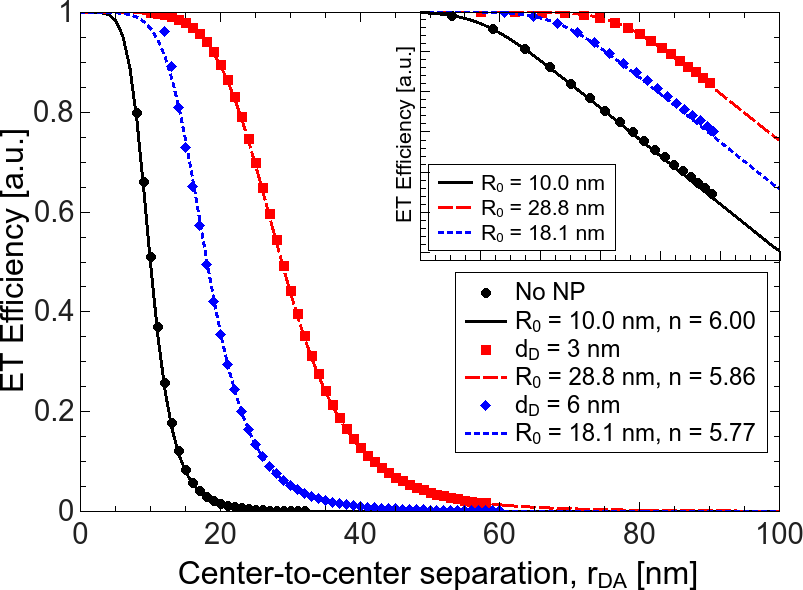}}\hspace{1cm}
	\subfloat[$a = 2.5$ nm, fixed acceptor distance\label{fig:06b}]{\includegraphics[height=0.26\textwidth]{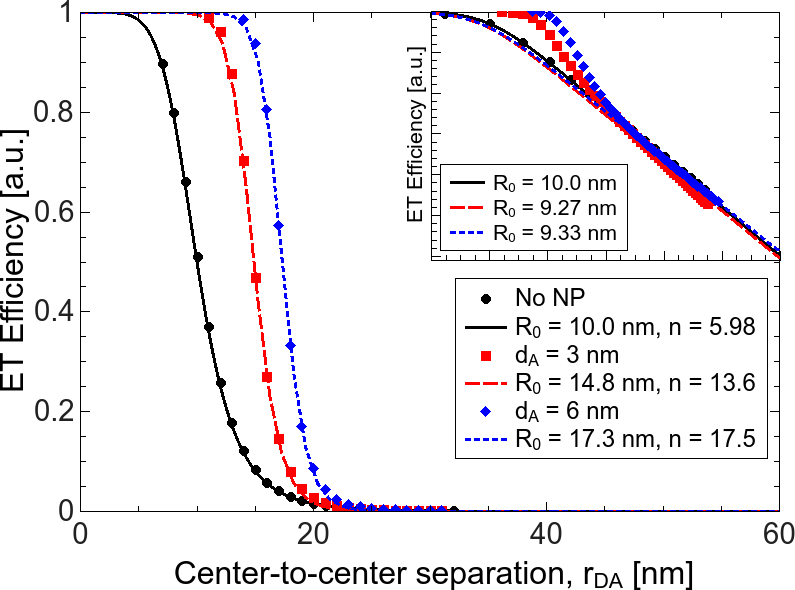}}\\
	\subfloat[$a = 5.0$ nm, fixed donor distance\label{fig:06c}]{\includegraphics[height=0.26\textwidth]{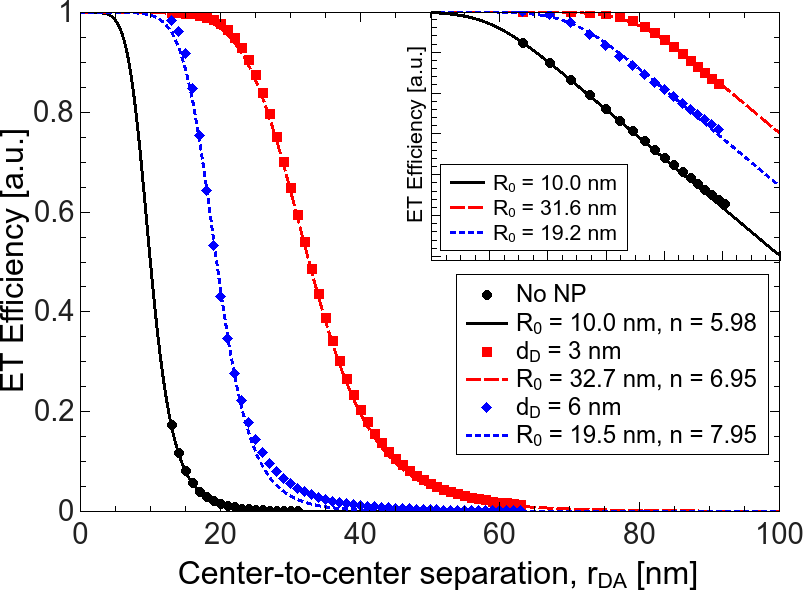}}\hspace{1cm}
	\subfloat[$a = 5.0$ nm, fixed acceptor distance\label{fig:06d}]{\includegraphics[height=0.26\textwidth]{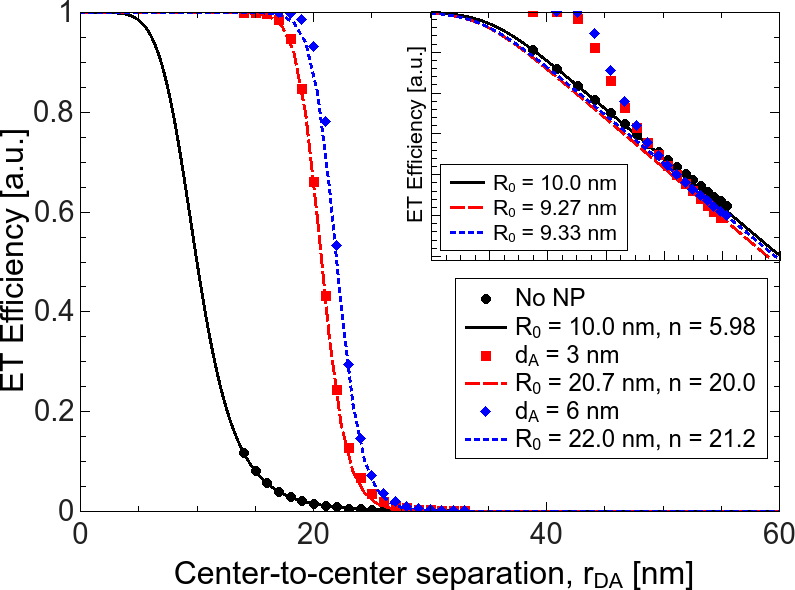}}\\
	\subfloat[$a = 10.0$ nm, fixed donor distance\label{fig:06e}]{\includegraphics[height=0.26\textwidth]{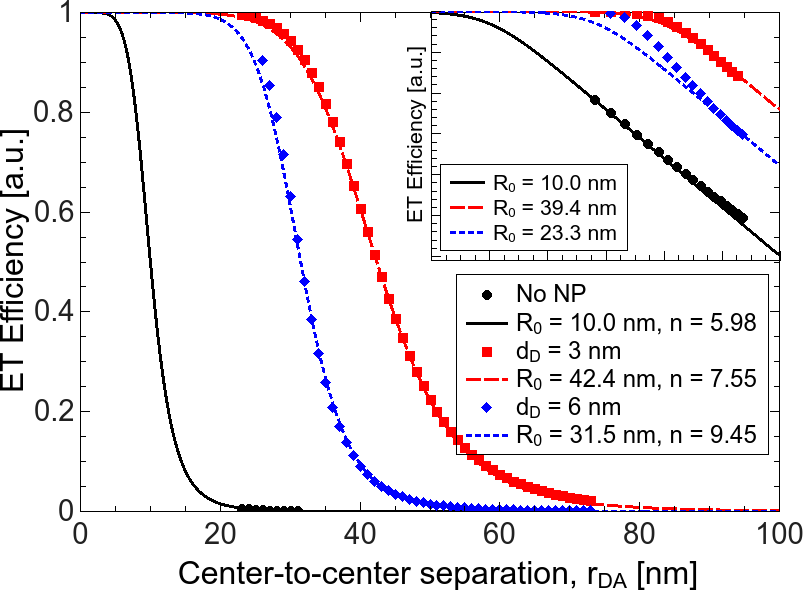}}\hspace{1cm}
	\subfloat[$a = 10.0$ nm, fixed acceptor distance\label{fig:06f}]{\includegraphics[height=0.26\textwidth]{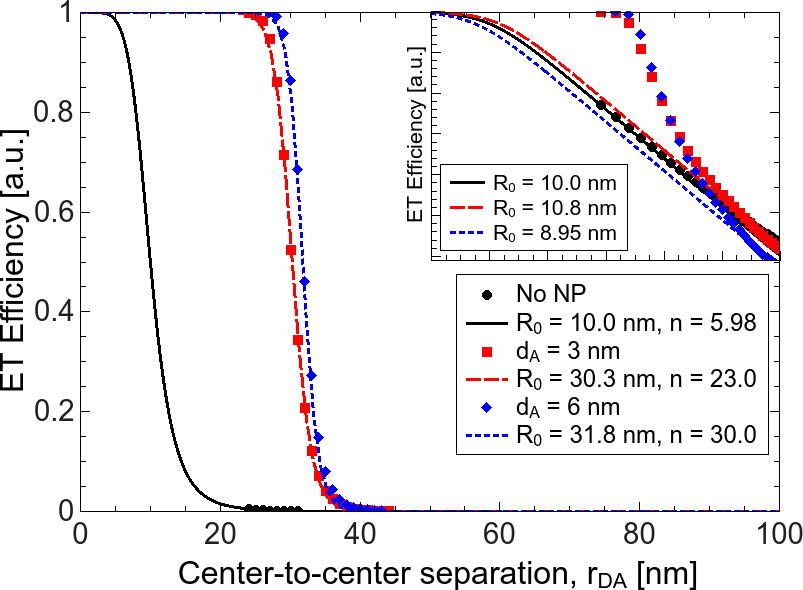}}\\
	\caption{(Color online) Energy transfer efficiency between a donor-acceptor pair on opposite sides of the Au NP, as a function of donor-acceptor separation, $r_\text{DA}$, when the donor-Au NP distance is kept fixed (a), (c), (b) and the acceptor-Au NP distance is kept fixed (b), (d), (f); The Au NP radius is $a = 2.5$ nm in (a) and (b), $a = 5.0$ nm in (c) and (d), and $a = 10$ nm in (e) and (f). The insets show the same data on a log-log scale.\label{fig:06}}
\end{figure*}

We first investigate the dependence of the energy transfer efficiency between a donor-acceptor dipole pair on their separation. The donor and acceptor are placed on opposite sides of a Au NP of various radii, forming a linear triad. Fig.~\ref{fig:06} shows the distance dependence of the energy transfer efficiency from donor to acceptor for the case when the distance from the donor (\ref{fig:06a}, \ref{fig:06c}, \ref{fig:06e}) or the acceptor (\ref{fig:06b}, \ref{fig:06d}, \ref{fig:06f}) to the Au NP is constant, while the position of the other member of the pair is varied.
Each panel of Fig.~\ref{fig:06} shows the energy transfer efficiency as a function of the donor-acceptor center-to-center separation, $r_\text{DA}$, for several Au NP radii, $a = 2.5$ nm for \ref{fig:06a} and \ref{fig:06b}, $a = 5.0$ nm for \ref{fig:06c} and \ref{fig:06d}, and $a = 10$ nm for \ref{fig:06e} and \ref{fig:06f}. Panel \ref{fig:06a} shows, in the main plot, the energy transfer efficiency between the donor and acceptor in the absence of the Au NP (black circles), when the donor is at $d_\text{D} = 3$ nm from the surface of the Au NP (red squares) and when the donor is at $d_\text{D} = 6$ nm from the surface of the Au NP (blue diamonds). Additionally, fits of the calculated data with a function of the form
\begin{equation}
	\eta(r_\text{DA}) = \frac{1}{1 + \left(\frac{r_\text{DA}}{R_0}\right)^n}
\end{equation}
are shown as lines for each data set. The data sets have been fitted to the curves in the region where $\eta(r_\text{DA}) = 50\%$ and then the fitting curves have been extended over the entire data range. The legend shows the fitting parameters $R_0$ and $n$. The inset in the panel shows the same data sets on a log-log scale, together with a different set of fitting curves. The procedure used to obtain this second set of curves has been to fix the value of the $n$ parameter to $n = 6$ (corresponding to FRET) and fit the data sets in the region where this power law holds, i.e.~at relatively large donor-acceptor separations. The legend in the inset shows the $R_0$ parameter extracted from these fits. Panel \ref{fig:06b} is analogous to \ref{fig:06a}, except that now the distance from the acceptor to the surface of the Au NP is kept fixed at $d_\text{A} = 3$ nm (red squares) and $d_\text{A} = 6$ nm (blue diamonds). Panels \ref{fig:06c}-\ref{fig:06f} are similar, except for different radii of the Au NP.

Beginning with panel \ref{fig:06a}, for which the donor-Au NP distance is kept fixed, it is evident that the energy transfer process between donor and acceptor follows a $r_\text{DA}^{-6}$ dependence on their separation, and is, hence, a FRET process. It is, however, a modified FRET process, since the characteristic distance or F{\"o}rster radius, $R_0$, takes on values dependent on the donor-Au NP distance and larger than the F{\"o}rster radius in the absence of the Au NP, which is $R_0 = 10.0$ nm. Thus, when the donor-Au NP distance is $d_\text{D} = 6$ nm, the F{\"o}rster radius increases to $R_0 = 18.1$ nm and at $d_\text{D} = 3$ nm, it is $R_0 = 28.8$ nm.

When considering panel \ref{fig:06b}, for which the acceptor-Au NP distance is kept fixed, the situation is quite different. The main plot shows an increased characteristic distance $R_0 = 14.8$ nm at $d_\text{A} = 3$ nm and $R_0 = 17.3$ nm at $d_\text{A} = 6$ nm, but the values of $n$ one obtains, $n = 17.5$ at $d_\text{A} = 3$ nm and $n = 13.6$ at $d_\text{A} = 6$ nm, are very different from $n = 6$, making it clear that this is no longer a FRET process, and the reduction of the energy transfer efficiency with distance is much more dramatic. We see in the inset that, were one to insist on a fit with a FRET model, that fit would be valid only for donor-acceptor separations larger than 20 nm. The $R_0$ extracted from this fit differs only negligibly from the free-space $R_0$ and is reduced, rather than enhanced.

The lessons to be drawn from the first two panels of Fig.~\ref{fig:06} are interesting: a fixed donor-Au NP separation leads to a FRET-type behavior of the energy transfer efficiency, with an increased characteristic distance $R_0$, which can be almost tripled. This suggests that such a system could be profitably used as a spectroscopic ruler: the FRET-type behavior assures a consistent distance dependence over large ranges, while the increased characteristic distance $R_0$ extends the functional range of the ruler. A fixed acceptor-Au NP system, on the other hand, is not suitable as a spectroscopic ruler because, even though the characteristic distance $R_0$ is increased, there is no consistent distance dependence of the energy transfer efficiency.

Considering now the second (\ref{fig:06c} and \ref{fig:06d}) and third (\ref{fig:06e} and \ref{fig:06f}) sets of panels in Fig.~\ref{fig:06}, the first things to notice is an increase in the characteristic distance, $R_0$, in the legends of the main panels, and a progressively more pronounced deviation of $n$ from the FRET value of $n = 6$. These are a result of increasing the radius of the Au NP from $a = 2.5$ nm to $a = 5.0$ nm, to $a = 10.0$ nm, with the effect that the range of donor-acceptor interaction is increased, and the higher-order multipoles of the Au NP contribute to this interaction. Secondly, the characteristic distances in the insets of each panel, extracted from a pure FRET fit, present an increase with the sphere size for a fixed donor-Au NP separation, but not for a fixed acceptor-Au NP separation. This happens because the FRET fit is valid only at large overall donor-acceptor separations, where the ET efficiency reverts to its value in the absence of the Au NP.

\subsubsection{Angular Position Dependence of the Energy Transfer Efficiencies}
\label{subsub:III.B.3}

\begin{figure*}[t]
	\subfloat[$a = 2.5\;\text{nm}, d_\text{D} = 3\;\text{nm}$\label{fig:07a}]{\centering\includegraphics[height=0.25\textwidth]{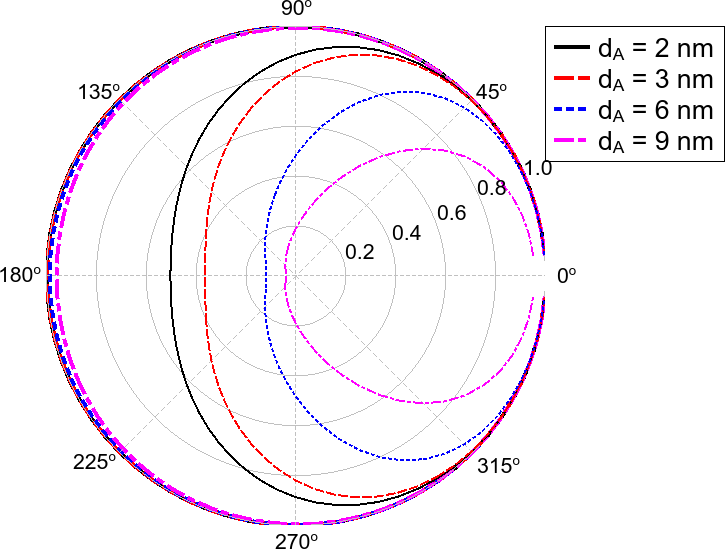}}\hfill
	\subfloat[$a = 5.0\;\text{nm}, d_\text{D} = 3\;\text{nm}$\label{fig:07b}]{\centering\includegraphics[height=0.25\textwidth]{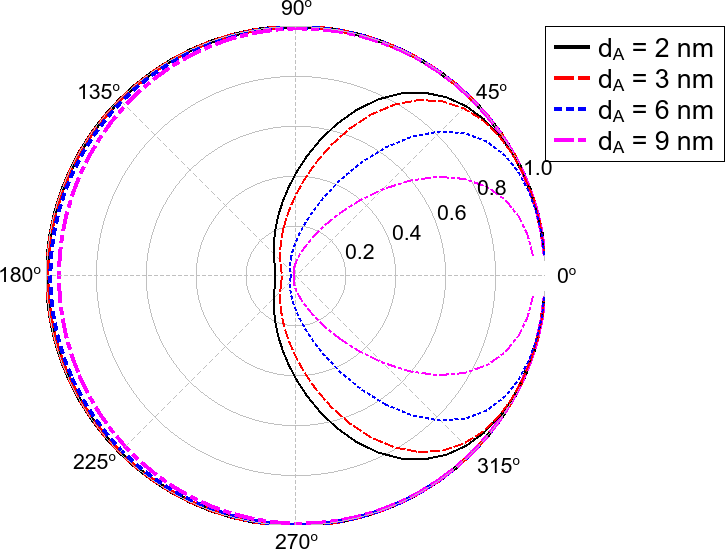}}\hfill
	\subfloat[$a = 10\;\text{nm}, d_\text{D} = 3\;\text{nm}$\label{fig:07c}]{\centering\includegraphics[height=0.25\textwidth]{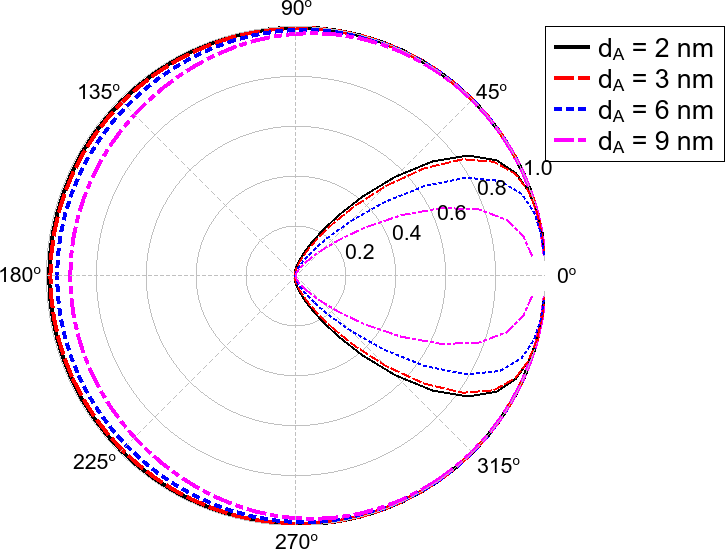}}\\
	\subfloat[$a = 2.5\;\text{nm}, d_\text{D} = 6\;\text{nm}$\label{fig:07d}]{\centering\includegraphics[height=0.25\textwidth]{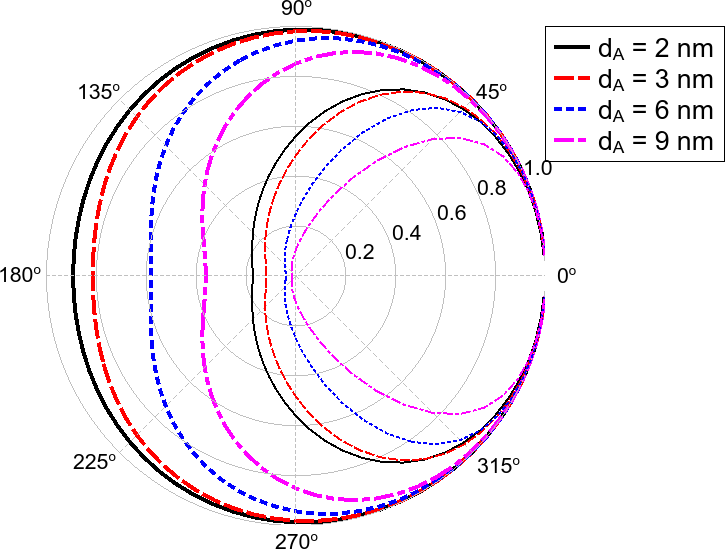}}\hfill
	\subfloat[$a = 5.0\;\text{nm}, d_\text{D} = 6\;\text{nm}$\label{fig:07e}]{\centering\includegraphics[height=0.25\textwidth]{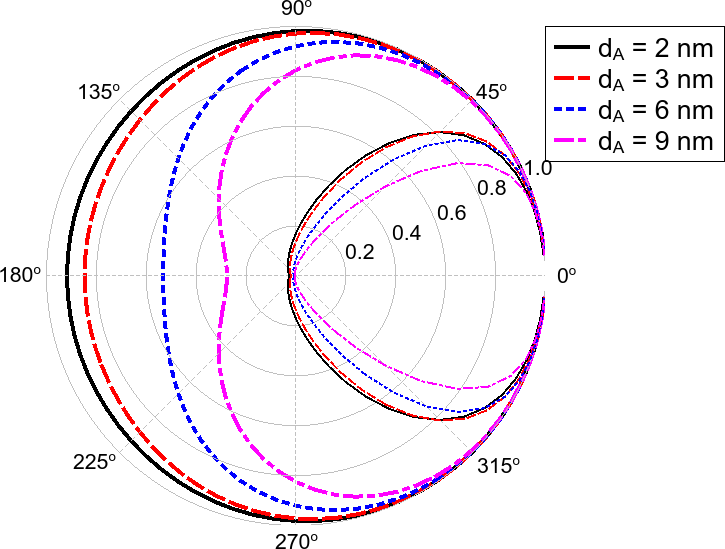}}\hfill
	\subfloat[$a = 10\;\text{nm}, d_\text{D} = 6\;\text{nm}$\label{fig:07f}]{\centering\includegraphics[height=0.25\textwidth]{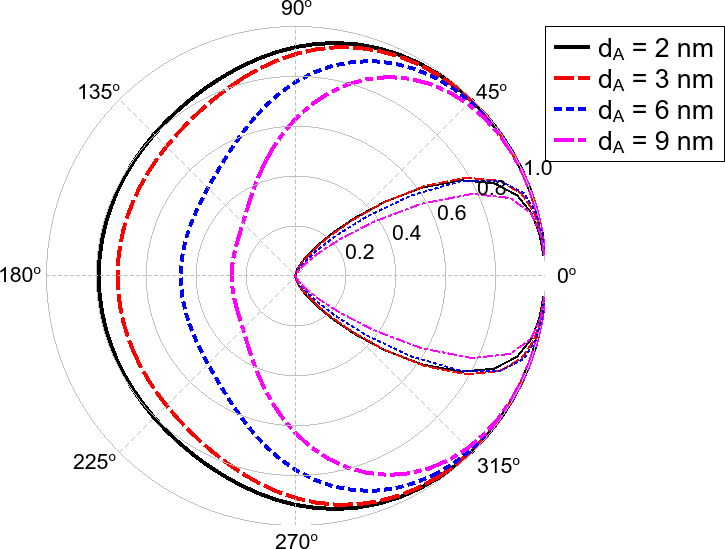}}\\
	\subfloat[$a = 2.5\;\text{nm}, d_\text{D} = 9\;\text{nm}$\label{fig:07g}]{\centering\includegraphics[height=0.25\textwidth]{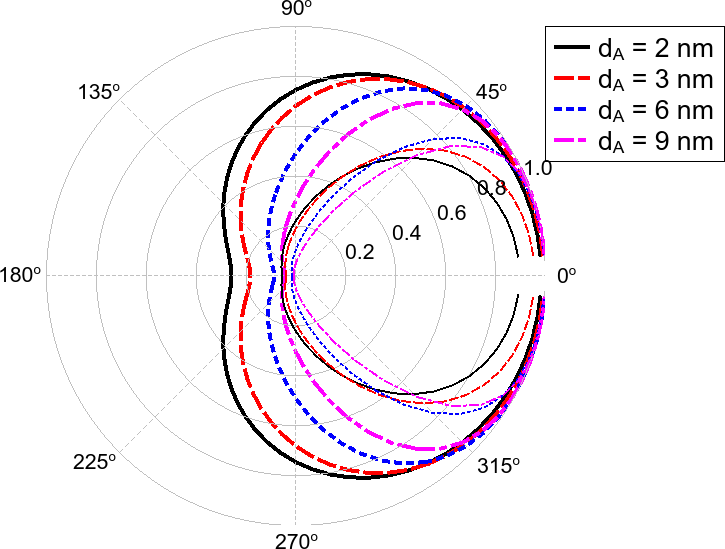}}\hfill
	\subfloat[$a = 5.0\;\text{nm}, d_\text{D} = 9\;\text{nm}$\label{fig:07h}]{\centering\includegraphics[height=0.25\textwidth]{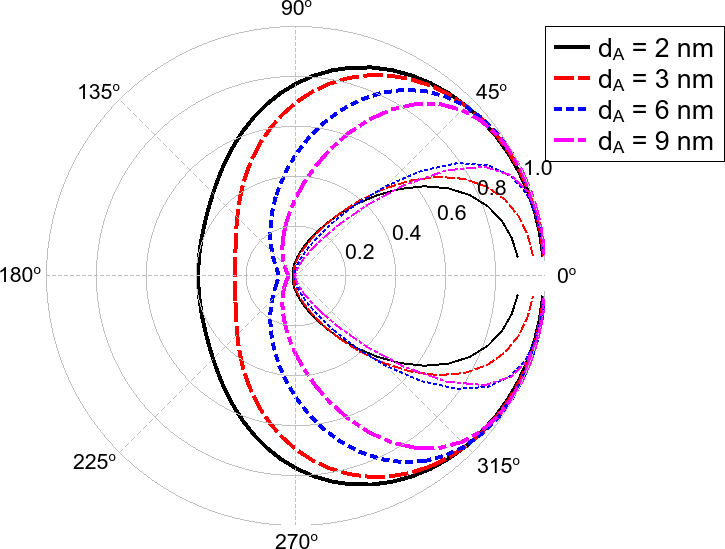}}\hfill
	\subfloat[$a = 10\;\text{nm}, d_\text{D} = 9\;\text{nm}$\label{fig:07i}]{\centering\includegraphics[height=0.25\textwidth]{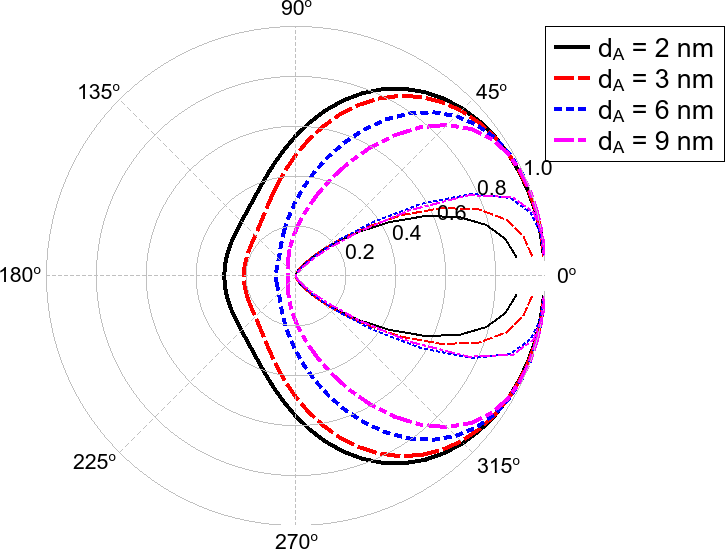}}\\
	\caption{(Color online) Polar plot of the ET efficiency between a donor located at $d_\text{D} = 3$ nm (top row), $d_\text{D} = 6$ nm (middle row) and $d_\text{D} = 9$ nm (bottom row) from the surface of the Au NP on the $x$-axis and an acceptor located at different distances, $d_\text{A}$, from the surface of a sphere of radius $a = 2.5$ nm (left column), $a = 5$ nm (middle column) and $a = 10$ nm (right column). The emission wavelength of the donor is $\lambda_\text{em} = 525$ nm. The orientations of the donor and acceptor dipoles are isotropic. The thin lines correspond to the same arrangements of donor and acceptor as the thick lines, except that they are calculated in the absence of the Au NP.\label{fig:07}}
\end{figure*}

In the previous subsection, the donor and acceptor were placed diametrically opposed around the sphere, i.e.~at antipodes, and the energy transfer efficiency was investigated as a function of their distance to the surface of the sphere alone. We now investigate the behavior of the energy transfer efficiency when the angular position of the donor and acceptor around the sphere is varied. Fig.~\ref{fig:07} shows polar plots of the ET efficiency for a donor-acceptor pair close to a Au NP of radius $a = 2.5\;\text{nm}$ (left column), $a = 5.0$ nm (middle column) and $a = 10$ nm (right column). The position of the donor is fixed on the $x$-axis at $d_\text{D} = 3$ nm (top row), $d_\text{D} = 6$ nm (middle row) and $d_\text{D} = 9$ nm (bottom row) from the surface of the Au NP, while several acceptor distances are considered (see legend in Fig.~\ref{fig:07}). The emission wavelength of the donor is $\lambda_\text{em} = 525\;\text{nm}$, and an average is performed over donor and acceptor dipole orientations.
The thin curves in each panel represent the ET efficiency in the absence of the Au NP, i.e.~in free-space, for exactly the same parameters.

For the first row of Fig.~\ref{fig:07}, comprising panels \ref{fig:07a}, \ref{fig:07b}, and \ref{fig:07c}, the distance of the donor from the surface of the Au NP of different sizes is $d_\text{D} = 3$ nm. In this case, the ET efficiency around the Au NP is very close to 100\%, for all three sphere sizes, and all acceptor distances to the surface of the Au NP, though it can be seen that the ET efficiency begins to decrease for the largest acceptor distance, $d_\text{A} = 9$ nm. This effect is more pronounced for the larger Au NP, for which the donor-acceptor separation is larger.

As the donor distance to the surface of the Au NP is increased, the angular dependence of the ET efficiency is different for Au NPs of various sizes. In the case of the smallest Au NP with a radius $a = 2.5$ nm (left column in Fig.~\ref{fig:07}), the angular dependence of the ET efficiency resembles the ET efficiency in the absence of the Au NP, albeit with larger overall values. Furthermore, the results obtained with this full method are practically indistinguisable from those obtained by treating the Au NP as a point dipole (data not shown). This strongly suggests that the FRET-type interaction we have seen being valid in the previous section on the distance dependence of the ET rate, also describes the angular dependence in the case of the smallest sphere, $a = 2.5$ nm.

For Au NP of larger sizes (the middle, $a = 5$ nm, and right columns, $a = 10$ nm, in Fig.~\ref{fig:07}), the angular behavior of the ET efficiency begins to deviate from a FRET-type. This is especially notable in the appearance of a bulge at an angle $\pi$, which contrasts with the dip observed in the free-space ET efficiency at the same angle. Moreover, when treating the Au NP as a point dipole, this bulge is absent, strongly suggesting that it is due to higher order multipoles of the Au NP. A final piece of evidence for this hypothesis is the fact that the bulge is less pronounced or dissappears when either the donor or acceptor is moved away from the surface of the Au NP and can no longer couple to these higher order modes.

\subsubsection{Spectral Overlap Dependence of the Energy Transfer Rates}
\label{subsub:III.B.4}

\begin{figure*}[t]
	\subfloat[$a = 2.5\;\text{nm}$\label{fig:08a}]{\centering\includegraphics[height=0.22\textwidth]{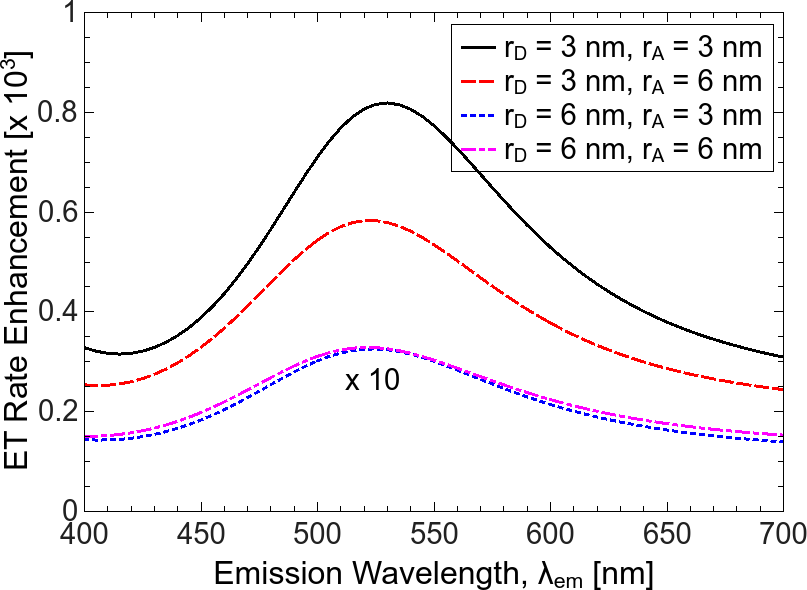}}\hfill
	\subfloat[$a = 5.0\;\text{nm}$\label{fig:08b}]{\centering\includegraphics[height=0.22\textwidth]{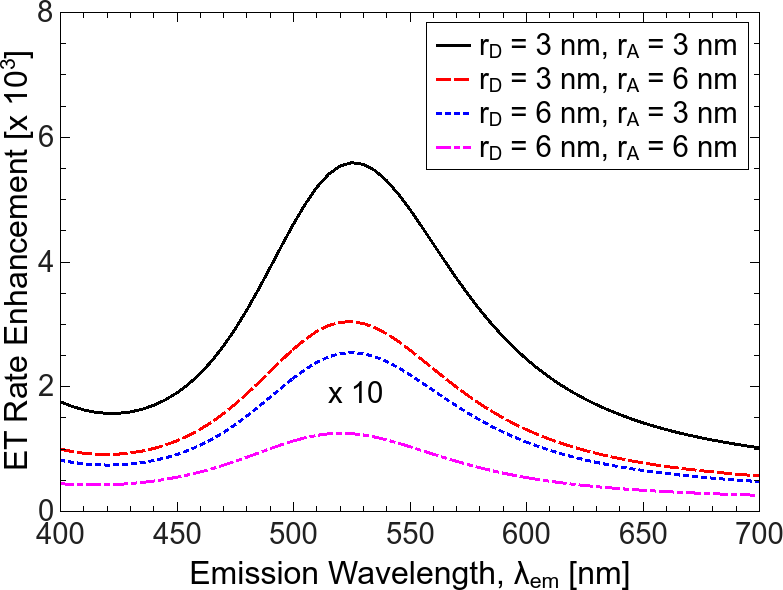}}\hfill
	\subfloat[$a = 7.5\;\text{nm}$\label{fig:08c}]{\centering\includegraphics[height=0.22\textwidth]{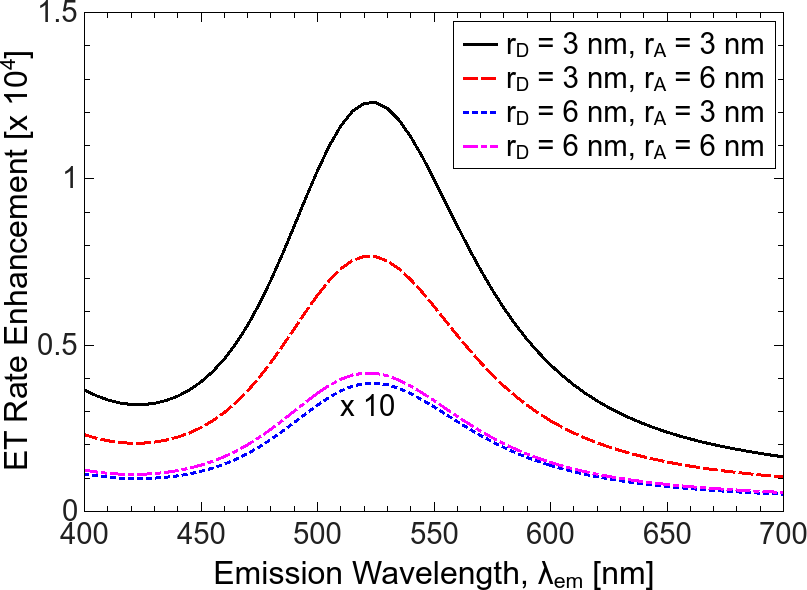}}\\
	\caption{(Color online) ET rate enhancement as a function of the central wavelength of the donor emission spectrum for several positions of the donor and acceptor and three sizes of the Au NP: (a) $a = 2.5$ nm, (b) $a = 5$ nm and (c) $a = 7.5$ nm.\label{fig:08}}
\end{figure*}

We now consider the dependence of the energy transfer rate and efficiency on the spectral overlap between the donor emission, acceptor absorption and the LSP peak of the Au NP. To this end we fix the position of the donor and acceptor at several distances from the surface of the Au NP, on opposite sides. We then consider gaussian distributions with a fixed width of 45 nm for the donor emission spectrum and we calculate the energy transfer rate as a function of the central wavelengths of these distributions. The results are shown in Fig.~\ref{fig:08} for three sizes of the Au NPs, $a = 5, 10, 15\;\text{nm}$.
As the figure shows, the ET rate enhancement peaks between 500 nm and 550 nm for all sizes. For the smallest size, $a = 2.5\;\text{nm}$ (panel \ref{fig:08a}), the ET enhancement peak is around 530 nm when both donor and acceptor are very close to the Au NP (3 nm) and it blue-shifts to 520 nm when the donor and acceptor are moved away from the Au NP. As the size of the Au NP is increased to $a = 5\;\text{nm}$ (Fig.~\ref{fig:08b}) and $a = 7.5\;\text{nm}$ (Fig.~\ref{fig:08c}), the ET rate enhancement peak is around 525 nm and is not noticeably dependent on the position of either the donor or acceptor. For the case of the smallest Au NP, Fig.~\ref{fig:08a} with $a = 2.5\;\text{nm}$, the LSP is around $\lambda = 500\;\text{nm}$. The maximum ET rate enhancement for this case, however, occurs between the LSP peak at 500 nm and the near-field scattering peak at 540 nm (see Fig.~\ref{fig:02} for the spectral shapes of the extinction and the near-field scattering efficiencies of Au NPs of various sizes). This reflects the influence of the near-field of the Au NP, which has evanescent components not visible in far-field measurements, but which contribute to ET close to the Au NP.

Another feature of the plots in Fig.~\ref{fig:08} is the broadening of the spectra with decreasing Au NP size. This is clearly seen in Fig.~\ref{fig:08} and is a consequence of the broadening of the LSP peak due to a surface scattering contribution in the dielectric function of Au spheres of sizes below 50 nm.\cite{Averitt1997}

\subsubsection{Au NP Size Dependence of the Energy Transfer Efficiencies}
\label{subsub:III.B.5}
\begin{figure*}[t]
	\subfloat[$d_\text{D} = 1\;\text{nm}$\label{fig:09a}]{\centering\includegraphics[width=0.49\textwidth]{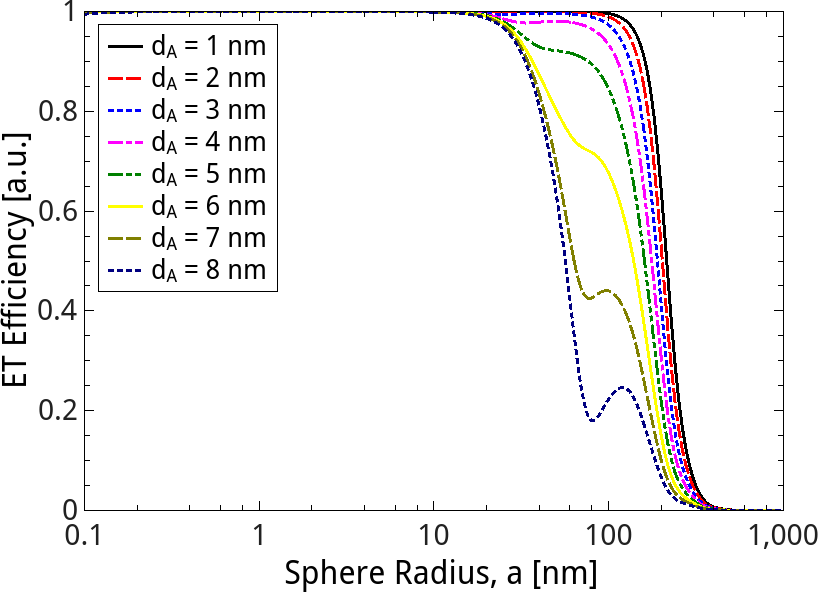}}\hfill
	\subfloat[$d_\text{D} = 3\;\text{nm}$\label{fig:09b}]{\centering\includegraphics[width=0.49\textwidth]{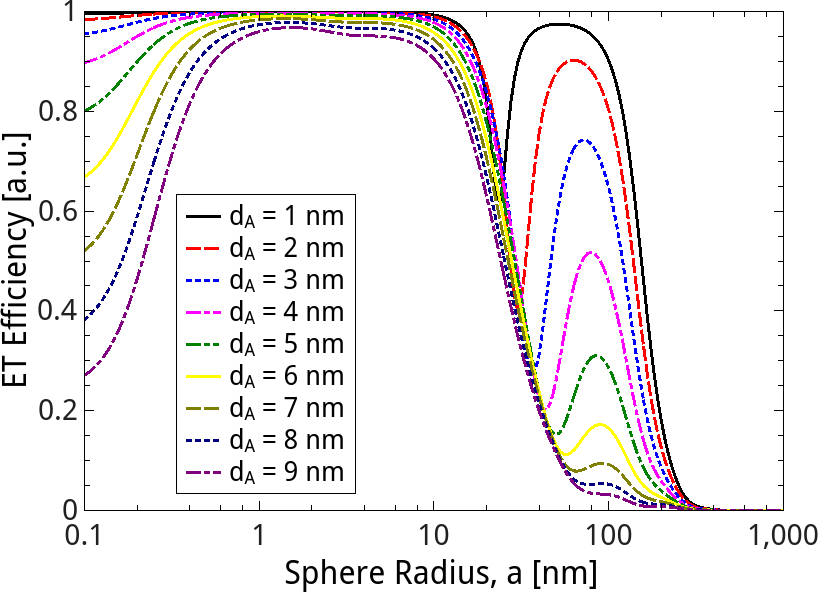}}\\
	\subfloat[$d_\text{D} = 5\;\text{nm}$\label{fig:09c}]{\centering\includegraphics[width=0.49\textwidth]{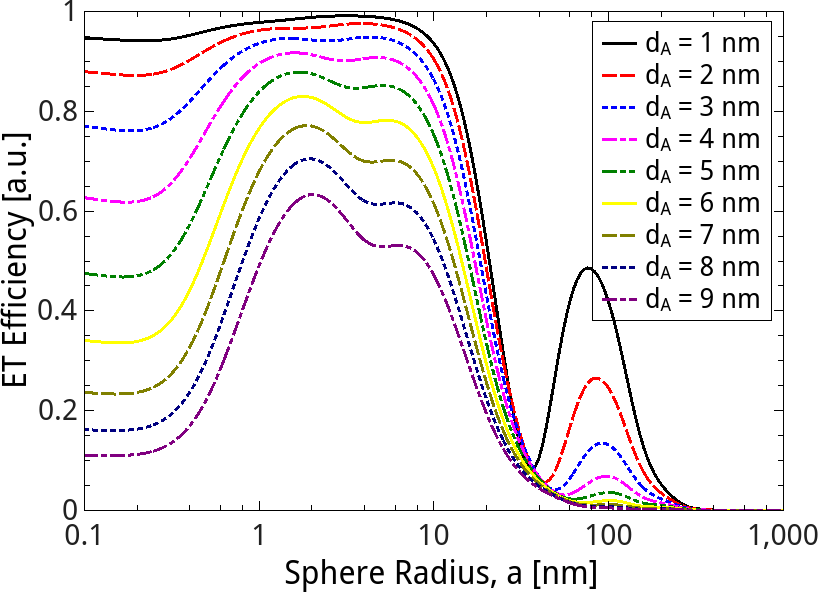}}\hfill
	\subfloat[$d_\text{D} = 7\;\text{nm}$\label{fig:09d}]{\centering\includegraphics[width=0.49\textwidth]{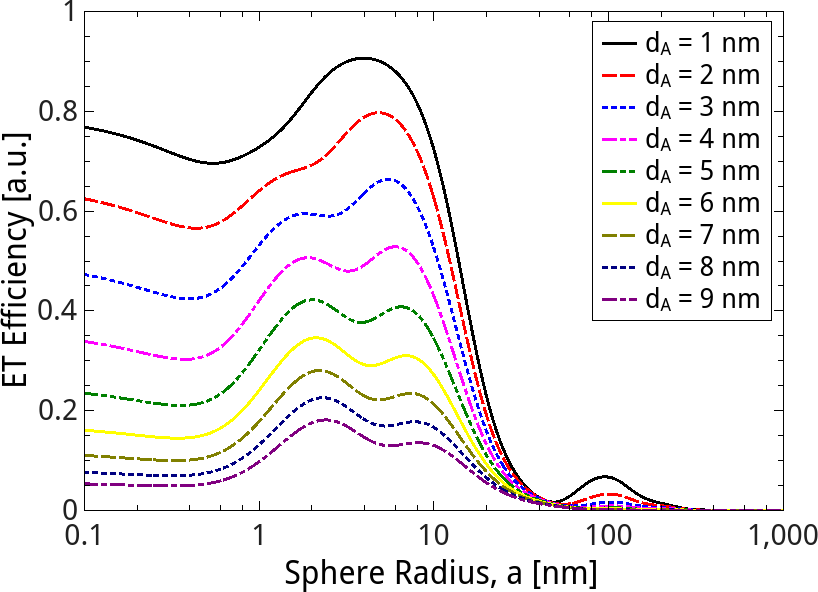}}\\
	\caption{(Color online) ET Efficiency as a function of the sphere radius, $a$, for several acceptor positions and four donor positions, (a) $d_\text{D} = 1\;\text{nm}$, (b) $d_\text{D} = 3\;\text{nm}$, (c) $d_\text{D} = 5\;\text{nm}$ and (d) $d_\text{D} = 7\;\text{nm}$.\label{fig:09}}
\end{figure*}

Finally, we investigate the dependence of the ET efficiency on the size of the Au NP. Fig.~\ref{fig:09} shows the sphere radius dependence of the ET efficiency for several distances of the donor from the surface of the Au NP, $d_\text{D} = 1$ nm (panel \ref{fig:09a}), $d_\text{D} = 3$ nm (panel \ref{fig:09b}), $d_\text{D} = 5$ nm (panel \ref{fig:09c}) and $d_\text{D} = 7$ nm (panel \ref{fig:09d}), and several distances of the acceptor from the surface of the Au NP ($d_\text{A} = 1$ nm to $d_\text{A} = 9$ nm).
The donor and acceptor are on opposite sides of the Au NP. When the donor distance to the surface of the Au NP is small, as in panel \ref{fig:09a}, the ET efficiency is almost 100\%, for sphere radii up to 100 nm, when the acceptor distance to the surface is also relatively small. As the acceptor distance to the surface is increased, panel \ref{fig:09a} shows the development of a peak in the ET efficiency around and above 100 nm. This extra peak is not present when the Au NP is treated in the dipole approximation; indeed, only when one takes into account multipole orders above the tenth, does the peak appear. It is therefore reasonable to attribute this peak to the higher order multipoles of the Au nanosphere. This is corroborated by the fact that the height of the peak decreases rapidly as the distance of either the donor or acceptor to the surface of the Au NP is increased. This suggests that the effect belongs to the near-field of the Au NP, which is in accordance with the effect being due to the higher multipoles of the Au nanosphere. Increasing the donor distance to the surface of the Au NP, as in panels \ref{fig:09b} to \ref{fig:09d}, the behaviour of the ET efficiency changes quite dramatically, especially for very small sphere radii. When the radius of the Au NP is significantly smaller than the distance of either the donor or acceptor to the surface of the Au NP, the ET efficiency can be considerably reduced. Our simulations show, in fact, that in the limit $a \to 0$, the ET efficiency in the presence of the Au NP is equal to the ET efficiency in free-space (data not shown).

\section{Summary and Conclusions}
\label{sec:IV}

In this contribution we have investigated the quenching of the emission of a quantum emitter placed close to Au NPs. We have employed a Green's tensor formalism, which we have first validated by simulating experimental data obtained in our lab, for monolayers of QDs on top of monolayers of Au NPs. We have found good agreement between this parameter-free simulation and the experimental measurements. Subsequently, we have employed the Green's tensor formalism to investigate the dependence of the quenching efficiency of a single QE close to a single Au NP on several parameters, such as the emission wavelength, intrinsic quantum yield of the QE, the size of the Au NP and the QE-Au NP distance.

We have considered the dependence of the quenching efficiency on the emission wavelength and intrinsic quantum yield of the QE together, since these two parameters are often modified in tandem in experiments. We have found that the characteristic distance of the QD-Au NP pair increases monotonically with the quantum yield of the emitter, whereas the dependence on the emission wavelength of the emitter has a rather pronounced peak located between the LSP wavelength of the Au NP and the near-field scattering efficiency maximum. When considering the dependence of the quenching efficiency -- or characteristic distance -- experimentally, it is therefore important to take into account both of these dependences, as not to do so can give artifacts, such as an apparent shift of the maximum quenching efficiency with the emission wavelength of the emitter.

We have also investigated the distance dependence of the quenching efficiency, mainly to ascertain under which conditions the FRET model for the QE -- Au NP interaction applies. Our simulations show that the FRET model provides a good approximation of the interaction, especially close to the LSP wavelength and for NP sizes below $5\,\text{nm}$. For larger spheres, not only does the exponent $n$ deviate from the value of 6 associated with the FRET model, but the fit becomes poorer, suggesting that the single power law used in the fit no longer accurately applies. 

We must mention here that our model does not reproduce the experimental results obtained by the Strousse group\cite{Yun2005,Jennings2006,Singh2010} for the energy transfer from a quantum emitter, a fluorescent dye in this case, to a Au NP with a diameter of approximately $D = 1.5$ nm. These authors have found that the distance dependence of the quenching efficiency is better described by the NSET model, with a $d^{-4}$ distance dependence. Evidently, more research is required to elucidate this discrepancy.

Finally, considering the behavior of the quenching efficiency as a function of the size of the Au NP, we have seen that the characteristic distance to the Au NP surface is determined by the extinction cross-section of the NP. Thus, when the size of the Au NP is much smaller than the emission wavelength of the QE, the extinction cross-section is dominated by absorption and the characteristic distance can be 2-3 times larger than the NP radius, following a power law dependence on the Au NP radius. In the size regime where scattering dominates the extinction cross-section, this distance becomes constant and independent on the Au NP radius. Quenching, in this latter case, occurs mainly very close to the NP surface and the QE emission can be expected to be enhanced and not quenched further away from the surface.

We have also investigated the energy transfer between a donor and acceptor. The energy transfer efficiency is much more sensitive to the donor-Au NP distance, than to the acceptor-Au NP distance. Furthermore, the distance dependence of the energy transfer efficiency for fixed donor-Au NP separations suggests that the donor-acceptor interaction mediated by the Au NP is essentially of F{\"o}rster type, following a $r^{-6}$ dependence on the donor-acceptor separation $r$ over a large distance range, but with an increased characteristic distance. This is not the case when the acceptor-Au NP separation is fixed. In this case the energy-transfer efficiency does not follow a F{\"o}rster dependence and it is influenced by the presence of the Au NP only for small donor-Au NP separations.

The positioning of the donor and acceptor around the sphere can also influence the energy transfer, especially for smaller Au NPs and donors and acceptors placed very close to the Au NP. For small Au NP sizes, the angular distribution of the ET efficiency is similar to that for a dipole pair, albeit with an increased characteristic distance. As the size of the Au NP is increased, however, there is a marked deviation from the dipole pair behavior at an angle of $\pi$. This can be attributed to the contribution of the higher order multipoles of the Au NP to the energy transfer process.

When investigating the dependence of the ET transfer efficiency between a donor-acceptor pair placed near the Au NP on the emission wavelength of the donor, we have shown that the largest ET rate does not occur at the LSP peak, but is red-shifted close to the peak in the near-field scattering efficiency. This can be attributed to the fact that, when the donor-acceptor pair is placed in the near-field of the Au NP, the evanescent components of the electric field can act as a strong channel for the ET process. This is more clearly apparent when one considers the ET rate enhancement dependence on the donor-acceptor distance to the Au NP. In this case a gradual blue-shift towards the LSP wavelength occurs when the distance is increased and, hence, the evanescent component contribution becomes negligible.

Finally, an interesting behaviour of the ET efficiency as a function of sphere radius has been obtained from our simulations. For relatively small distances of the donor and acceptor to the surface of the Au NP, in addition to the large ET efficiency obtained for relatively small radii, there is an extra peak in the ET efficiency when the radius is around 100 nm. We have shown that this peak is a consequence of the higher multipole orders of the Au nanosphere contributing to the energy transfer process, for large enough spheres.

In this contribution we have undertaken a thorough investigation of the effects of a Au NP of subwavelength sizes on the decay and energy transfer processes of quantum systems placed in its vicinity. The increased characteristic distance provided by the Au NP can have interesting applications in improved sensing and as an extended spectroscopic ruler.

\begin{acknowledgments}
This work was supported by the Science Foundation Ireland under grant No. 10/IN.1/12975.
\end{acknowledgments}

\bibliographystyle{prsty}

\end{document}